\documentclass[twocolumn]{aastex631}

\begin{document}

\title{Challenging the AGN scenario for {\it JWST}/NIRSpec LRD and non-LRD broad H$\alpha$ emitters in light of non-detection of NIRCam photometric variability and X-ray}

\correspondingauthor{Mitsuru Kokubo}
\email{mitsuru.kokubo@nao.ac.jp}

\author[0000-0001-6402-1415]{Mitsuru Kokubo}
\altaffiliation{NAOJ Fellow}
\affiliation{National Astronomical Observatory of Japan, National Institutes of Natural Sciences, 2-21-1 Osawa, Mitaka, Tokyo 181-8588, Japan}

\author[0000-0002-6047-430X]{Yuichi Harikane}
\affiliation{Institute for Cosmic Ray Research, The University of Tokyo, 5-1-5 Kashiwanoha, Kashiwa, Chiba 277-8582, Japan}




\begin{abstract}

{\it JWST} has uncovered a substantial population of high-$z$ ($z \gtrsim 4$) galaxies exhibiting broad H$\alpha$ emission line with a Full Width at Half Maximum exceeding $1,000$~km~s$^{-1}$.
This population consists of a well-known subset called Little Red Dots (LRDs) and the remaining non-LRDs.
If all of these broad H$\alpha$ emitters were attributed to type $1-1.9$ Active Galactic Nuclei (AGNs), it would imply a significantly higher number density of low-luminosity AGNs than extrapolated from that of more luminous AGNs.
Here, we have examined the rest-frame ultraviolet (UV)-optical flux variability of three LRD and two non-LRD broad H$\alpha$ emitters using multi-epoch, multi-band {\it JWST}/NIRCam imaging data.
The rest-frame temporal sampling interval of the NIRCam data ($\sim 400-500$~days$/(1+z)$) is comparable to typical variability timescales of AGNs with black hole (BH) masses of $M_{\text{BH}} \sim 10^{7}~M_{\odot}$; thus, the flux variations should be detectable if AGNs were present.
However, no measurable flux variation over the rest-frame wavelength range of $\lambda_{\text{rest}} \sim 1,500-9,000$\AA\ has been detected, placing stringent upper limits on the variability amplitudes.
This result, combined with the X-ray faintness confirmed by the ultra-deep {\it Chandra} data, indicates that, under the AGN scenario, we need to postulate peculiar Compton-thick broad-line AGNs with either ($a$) an intrinsically non-variable AGN disk continuum, ($b$) a host galaxy-dominated continuum, or ($c$) scattering-dominated AGN emission.
Alternatively, ($d$) they could be non-AGNs where the broad-line emission originates from unusually fast and dense/low-metallicity star-formation-driven outflows or inelastic Raman scattering of stellar UV continua by neutral hydrogen atoms.

\end{abstract}



\section{Introduction}
\label{sec:intro}

Recent spectroscopic surveys by the James Webb Space Telescope ({\it JWST}) reveal a large number of high-$z$ galaxies at $z \gtrsim 4$ that exhibit broad ($\gtrsim 1,000$~km~s${}^{-1}$) hydrogen Balmer emission lines \citep[e.g.,][]{mai23,koc23,har23,kok23,kil23,mat24,gre24,koc24,aki24}.
Some of them have been discovered serendipitously throughout {\it JWST}/NIRSpec or NIRCam Grism spectroscopic surveys for galaxies \citep{mai23,koc23,fuj23,har23,mat24,mai24}, and the others have been identified through NIRSpec spectroscopic follow-ups of photometrically-selected AGN candidates that are collectively referred to as ``Little Red Dots'' (LRDs) photometrically characterized by its ultracompact morphology, unusually red rest-frame optical colors, and blue UV excess \citep[$v$-shaped' spectral energy distribution;][]{kil23,kok23,bar24,gre24,koc24,wan24,aki24}.

Except for a few exceptional cases, the broad emission component is observed only in H$\alpha$ (and sometimes in H$\beta$), whereas the other broad lines are not observed either due to dust reddening, low signal-to-noise ratio of the spectra, or intrinsic weakness of the emission lines \citep[e.g.,][]{kil23,koc24}.
The absence of the broad component in the strong [\ion{O}{3}] forbidden emission lines suggests that the broad line-emitting region cannot be low-density ionized plasma observed in normal galaxies \citep[e.g.,][]{gre24,mai24}.
Hereafter, we refer to all {\it JWST} objects exhibiting broad emission lines as {\it JWST} broad H$\alpha$ emitters, and among these, objects photometrically identified as LRDs (see Section~\ref{sec:sample_selection}) are termed LRD broad H$\alpha$ emitters, while those that do not exhibit LRD-like observational properties are referred to as non-LRDs broad H$\alpha$ emitters.

The broad H$\alpha$ emitters are not necessarily LRDs, and conversely, LRDs are not necessarily broad H$\alpha$ emitters.
Photometrically-selected LRDs are selected using a variety of criteria and thus constitute an inhomogeneous sample. The spectroscopically identified broad-line LRDs represent only a modest fraction of photometrically-selected LRDs, and the relationships among broad-line non-LRDs, broad-line LRDs, and non-broad-line (or narrow-line) LRDs remain a matter of ongoing discussion \citep{mat24,gre24,per24,koc24}.
The non-broad line LRDs may possibly be dusty compact star-forming galaxies \citep{wil24,per24,aki24}, whereas the broad H$\alpha$ emission lines cannot be attributable to normal stellar activities.
Thus, it is widely accepted in the literature that Active Galactic Nuclei (AGNs) somehow contribute to the observed properties of the broad H$\alpha$ emitters.
In this paper, we primarily focus on the broad-line population, unless otherwise stated.

Under this AGN scenario, the broad H$\alpha$ emission can be attributed to the AGN broad line region (BLR) emission.
The diversity of the contamination from the host galaxy light, dust extinction, and electron/dust-scattered AGN emission is invoked to explain the variety of the broad-band SED shape of the LRD and non-LRD broad H$\alpha$ emitters \citep{ono23,lab23,nob23,kil23,bar24,gre24,per24,wan24}.
For example, \cite{gre24} suggest that, while the rest-frame optical continuum of LRDs can be interpreted as a mildly-obscured broad-line AGN continuum, the origin of their UV emission can be more complicated; the UV continuum emission may be explained by the scattered AGN light when the ratio of the observed $L_{3000\text{\AA}}$ to the expected intrinsic $L_{3000\text{\AA}}$ is $\sim$ 1\%–3\% percent, while a star formation contribution to the UV needs to be invoked when the ratio exceeds 10\% \citep[see also][]{koc24}.
Given the significant flux contribution from the host galaxy, the detection of the broad H$\alpha$ is suggested to be the most powerful and reliable way to identify unobscured/mildly-obscured, low-luminosity, low-mass AGNs \citep{ono23,koc23,lab23,kil23,mat24}.

However, if all of the members in this abundant population of high-$z$ broad H$\alpha$ emitters were truly faint AGNs, then it would severely contradict the current understanding of the AGN population in several aspects.
The spatial density of the {\it JWST} broad H$\alpha$ emitters is $\gtrsim 10$ greater than the expectation extrapolated from the UV-optical and X-ray observations of higher luminosity AGNs \citep{gia19,har23,gre24,aki24}.
This means that the AGN occupation fraction in low-mass galaxies is extremely high, exceeding $\sim$ 5\% \citep{har23}.
The black hole (BH) mass to host galaxy stellar mass ratios $M_{\text{BH}}/M_{*}$ of the broad H$\alpha$ emitters pose another problem; the BH masses in these objects tend to be overmassive by a factor of $10-100$ compared to the $M_{\text{BH}}-M_{*}$ relationship in the local universe (\citealt{har23,pac23,dur24}, but see also \citealt{li24}).
The high spatial density of the broad H$\alpha$ emitter at $\gtrsim 4$, if confirmed as bona-fide AGNs with overmassive $M_{\text{BH}}$, is hard to explain without invoking a peculiar population of rapidly-spinning (radiatively-efficient) low-mass AGNs in the high-$z$ universe or very massive seed BH formation channels \citep[e.g.,][]{ina24b,jeo24}.
Moreover, the {\it JWST} broad H$\alpha$ emitters at $z \gtrsim 4$ would have produced a huge amount of X-ray photons if all of them are assumed to be AGNs, which may contradict the measurement of the spatially-unresolved X-ray background radiation \citep{pad23} \citep[but see also][]{yue24,ana24,mai24,mad24}.
It should be noted that so far no broad H$\alpha$ emitters (except for a few very bright LRDs) are confirmed to exhibit AGN hard X-ray emission, with neither individual detections nor stacking analyses yielding any signal, thus placing very stringent upper limits on their X-ray luminosities \citep[e.g.,][]{koc24,yue24,ana24,mai24,aki24}.

To avoid introducing such a peculiar population of low-mass AGNs, it is worthwhile to investigate alternative scenarios other than the AGN scenario to explain the observed properties of the {\it JWST} broad H$\alpha$ emitters.
Currently, the {\it JWST} broad H$\alpha$ emitters are regarded as AGN candidates primarily through the presence of the broad H$\alpha$; thus, they are strictly speaking type 1.9 AGNs.
However, not only the AGN broad line region but also some energetic stellar eruptions/explosions, such as massive stars' outflow, eruptions of Luminous Blue Variables (LBVs), Type IIn supernovae (SNe), and Tidai Disruption Events (TDEs) would also give rise to broad and long-lasting optical emission lines \citep[e.g.,][]{izo07,simmonds16,kok19,gre20,kok22,ina24a,gus24,juo24,mai24,wan24}.

Moreover, the inelastic scattering of UV photons by neutral hydrogen atoms might produce broad emission features over the optical-NIR wavelength range, which could mimic the AGN broad emission line \citep{kok24}.
Such atomic hydrogen scattering processes might be more common in higher-$z$ star-forming galaxies where abundant primordial neutral atomic hydrogen gas is present \citep[e.g.,][]{hei24}.
Interestingly, two out of the 20 broad H$\alpha$ emitters in the \cite{mat24} sample and three out of the 15 LRD broad H$\alpha$ emitters in the \cite{koc24} sample exhibit blue-shifted absorption in their Balmer lines that is rarely seen in lower-$z$ AGNs/galaxies \citep[see also][]{mai24}.
The Balmer absorption requires a very high neutral hydrogen gas column of $\sim 10^{19}~\text{cm}^{-2}$, and such a high detection rate of rare H$\alpha$ absorption features indicates that some physical processes uncommon in the lower-$z$ Universe are ongoing in these high-$z$ broad H$\alpha$ objects.

Therefore, it is essential to probe AGN signatures other than the broad H$\alpha$, such as broad \ion{C}{4} emission line and X-ray emission, to evaluate the contaminations from non-AGN H$\alpha$ broad-line objects and estimate the true volumetric density of the low-luminosity AGN population.
Unfortunately, most of the {\it JWST} broad H$\alpha$ emitters are red, which could be due to the mild dust extinction in the AGN scenario, and it would not be easy to detect the broad \ion{C}{4} emission line.
Also, the X-ray detection may be impractical given the intrinsic faintness \citep[e.g.,][]{bog24,koc24,ana24,mai24}.

In this work, we focus on the temporal variability of the AGN rest-frame UV-optical continuum, which is a ubiquitous property of the AGN accretion disk emission \citep[e.g.,][]{ulr97,ses07,kel09,mac10,mac12}.
Since the AGN variability amplitude is fairly large, especially in low-mass/luminosity AGNs ($\sim 0.1$~mag on a few month timescale), the AGN variability detection serves as a powerful way to identify unobscured/mildly-obscured AGNs even when they are buried under the host galaxy light \citep[e.g.,][]{kim20,bur23}.
If the {\it JWST} broad H$\alpha$ emitters are truly AGNs, the accretion disk continuum (at least at the wavelengths around and longer than the H$\alpha$ emission line) must be contributing to the observed continuum emission to some extent, and its flux variations must be observable even if they are mildly dust obscured.

It has been shown that the excellent sharpness and stability of the point spread function (PSF) of the Hubble Space Telescope ({\it HST}) imaging enable sensitive searches for the photometric variability of faint AGNs even up to $z \sim 6-7$ \citep[e.g.,][]{obr24,hay24}, and the {\it JWST} imaging will do a better job of detecting the variability \citep[e.g.,][]{mai24,jha24,dec24}.
By using the archival public {\it JWST}/NIRCam multi-band multi-epoch photometry data obtained in 2022 and 2023 in the Abell~2744 field, we investigate the flux variations of five broad H$\alpha$ emitters (two non-LRDs and three LRDs) in the sample of \cite{har23} and \cite{gre24} over the wavelength range of $\lambda_{\text{obs}} \sim 1-5~\mu$m, aiming at verifying/falsifying the AGN scenario for these broad H$\alpha$ emitters.

In Section~\ref{sec:data}, physical properties of the five {\it JWST} broad H$\alpha$ emitters studied in this work and NIRCam image processing are described.
The X-ray luminosity upper limits for the five objects obtained from the {\it Chandra} archival data are also provided in this section.
In Section~\ref{sec:analyses}, a search for the photometric variability in the NIRCam data is conducted, and upper limits on the variability amplitude in the five objects are obtained.
In Section~\ref{sec:discussion}, the implications of the non-detection of the X-ray and rest-frame optical photometric variability are discussed. 
The lack of variability in the five objects is contrasted with the variability observed in known AGNs, leading to the rejection of the standard AGN scenario for the broad H$\alpha$ emitters.
We consider several non-standard AGN and non-AGN models for the broad H$\alpha$ emitters, possibly explaining the non-detections of the photometric variability and X-ray simultaneously.
Finally, we summarize our conclusions and future prospects in Section~\ref{sec:conclusions}. 
We assume the flat $\Lambda$CDM cosmology with $H_{0} = 70~\text{km}~\text{s}^{-1}~\text{Mpc}^{-1}$, $\Omega_{\text{m}}=0.3$, and $\Omega_{\Lambda}=0.7$ when necessary.

\section{Data}
\label{sec:data}

\begin{figure*}[tbp]
\center{
\includegraphics[clip, trim=0.0cm 0.0cm 0.0cm 0.0cm, width=6.8in]{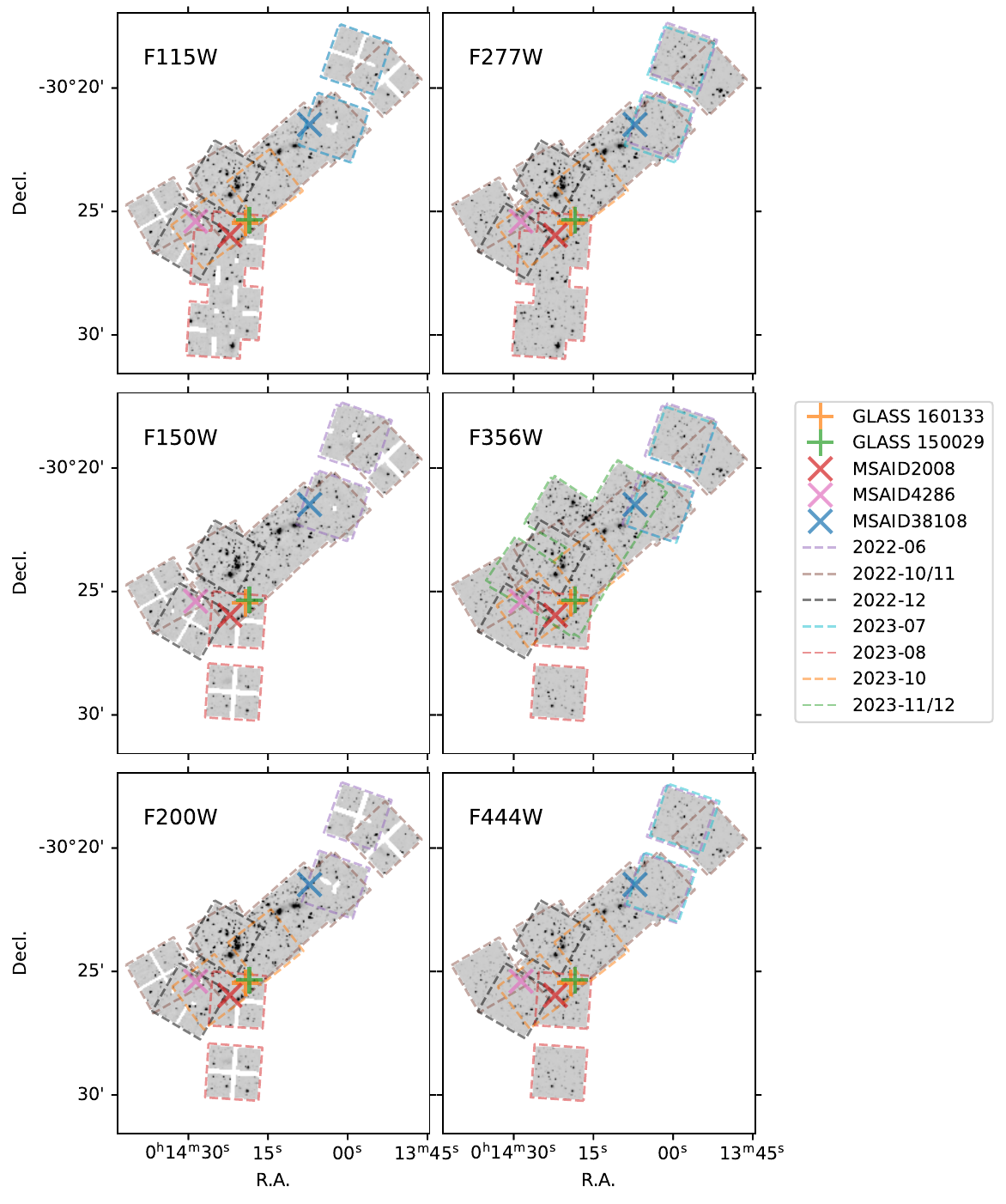}
}
 \caption{The NIRCam all-epoch mosaic images of the Abell~2744 field created from the multi-band multi-epoch data used in this study (see Section~\ref{sec:nircam_image_processing} and Table~\ref{tbl:summary_data}). Dashed lines in each panel denote the image footprints of the corresponding band, color-coded by the observation dates. The positions of the {\it JWST} broad H$\alpha$ emitters analyzed in this study are indicated by the $+$ and $\times$ symbols (\citealt{har23}'s non-LRD broad H$\alpha$ emitters; and \citealt{gre24}'s LRD broad H$\alpha$ emitters, respectively).
 }
 \label{fig:abell2744}
\end{figure*}

\begin{deluxetable*}{lllr}
\tablecolumns{4}
\tablecaption{Observation dates of the {\it JWST}/NIRCam wide-band data used in this work, and corresponding `epochs' for each of the {\it JWST} broad H$\alpha$ emitters (see Appendix~\ref{sec:log_of_observation} for details). \label{tbl:summary_data}}
\tablehead{
  \colhead{Date} & 
  \colhead{Bands} &
  \colhead{Program~ID} &
  \colhead{Object name (epoch)} \\
  \colhead{(YYYY-MM)} & 
  \colhead{} & 
  \colhead{} & 
  \colhead{}
  }
\startdata
  2022-06    & F115W F150W F200W F277W F356W F444W & 1324                  & MSAID38108   (epoch~1) \\\hline
  2022-10/11 & F115W F150W F200W F277W F356W F444W & 1324, 2561, 2756      & GLASS~160133 (epoch~1) \\
             &                                          &                  & GLASS~150029 (epoch~1) \\
             &                                          &                  & MSAID2008    (epoch~1) \\
             &                                          &                  & MSAID4286    (epoch~1) \\
             &                                          &                  & MSAID38108   (epoch~2) \\\hline
  2022-12    & F115W F150W F200W F277W F356W F444W & 2756                  & MSAID4286    (epoch~2) \\\hline
  2023-07    & F115W \hspace{62pt} F277W F356W F444W & 1324                & MSAID38108   (epoch~3) \\\hline
  2023-08    & F115W F150W F200W F277W F356W F444W & 2561                  & GLASS~160133 (epoch~2) \\
             &                                          &                  & GLASS~150029 (epoch~2) \\
             &                                          &                  & MSAID2008    (epoch~2) \\\hline
  2023-10    & F115W \hspace{30pt} F200W F277W F356W F444W & 3990          & MSAID2008    (epoch~3) \\
             &                                          &                  & MSAID4286    (epoch~3) \\\hline
  2023-11/12 & \hspace{127pt} F356W                     & 3516             & GLASS~160133 (epoch~3) \\
             &                                          &                  & GLASS~150029 (epoch~3) \\
             &                                          &                  & MSAID2008    (epoch~4) \\
             &                                          &                  & MSAID4286    (epoch~4) \\
             &                                          &                  & MSAID38108   (epoch~4) \\
\enddata
\tablenotetext{}{Program ID is the identifier for the JWST Early Release Science (ERS), General Observers (GO), and Director’s Discretionary (DD) programs: ERS 1324 (GLASS; PI: T. L. Treu), GO 2561 (UNCOVER; PI: I. Labbe), DD 2756 (PI: W. Chen), GO 3516 (PI: J. Matthee), and GO 3990 (BEACON; PI: T. Morishita).}
\end{deluxetable*}

\begin{figure*}[tbp]
\center{
\includegraphics[clip, trim=0.0cm 0.0cm 0.0cm 0.0cm, width=6.4in]{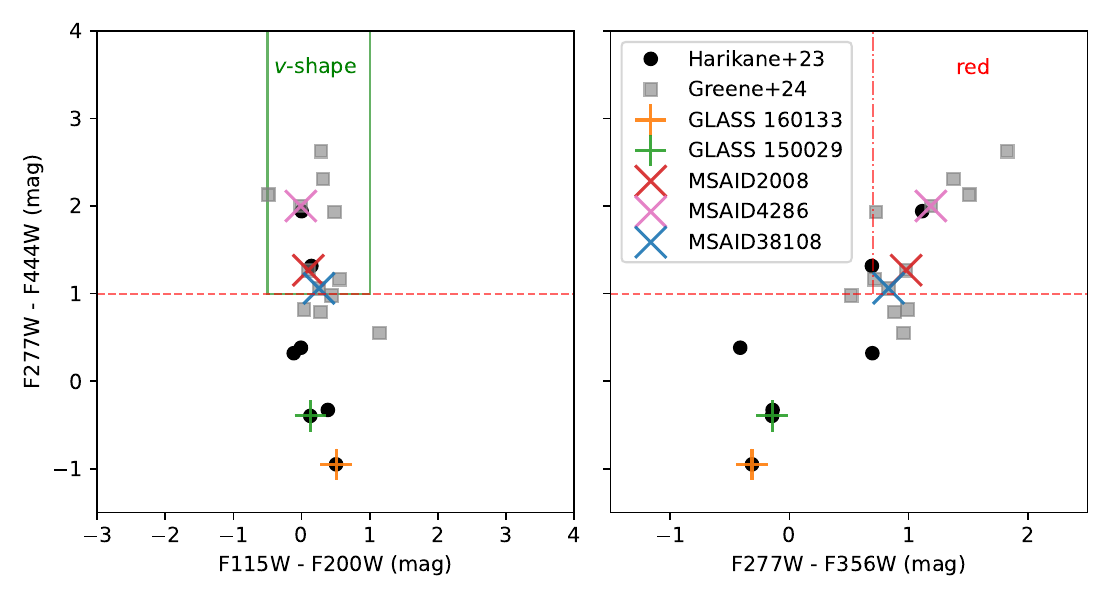}
}
 \caption{
 NIRCam $\mathrm{F115W}-\mathrm{F200W}$ versus $\mathrm{F277W}-\mathrm{F444W}$ and $\mathrm{F277W}-\mathrm{F356W}$ versus $\mathrm{F277W}-\mathrm{F444W}$ color–color diagrams of LRD and non-LRD broad H$\alpha$ emitters from \citet{har23} and \citet{gre24}. The photometry was taken from the CEERS and UNCOVER ``SUPER'' photometric catalogs \citep{wri24,sue24}. The broad H$\alpha$ emitters analyzed in this study are indicated by the $+$ and $\times$ symbols (\citealt{har23}'s non-LRD broad H$\alpha$ emitters; and \citealt{gre24}'s LRD broad H$\alpha$ emitters, respectively). The solid green line and the red dashed and dash-dotted lines represent the photometric LRD selection criteria defined by \citet{gre24}: the green line marks the $v$-shaped SED selection window, and the red dashed line ($\text{F277W}-\text{F444W} > 1$) and dash-dotted line ($\text{F277W}-\text{F356W} > 0.7$) indicate the selection criteria for the red AGN continua.
 }
 \label{fig:colorcolorspace}
\end{figure*}

\begin{figure}[tbp]
\center{
\includegraphics[clip, trim=0.0cm 0.0cm 0.8cm 0.0cm, width=3.4in]{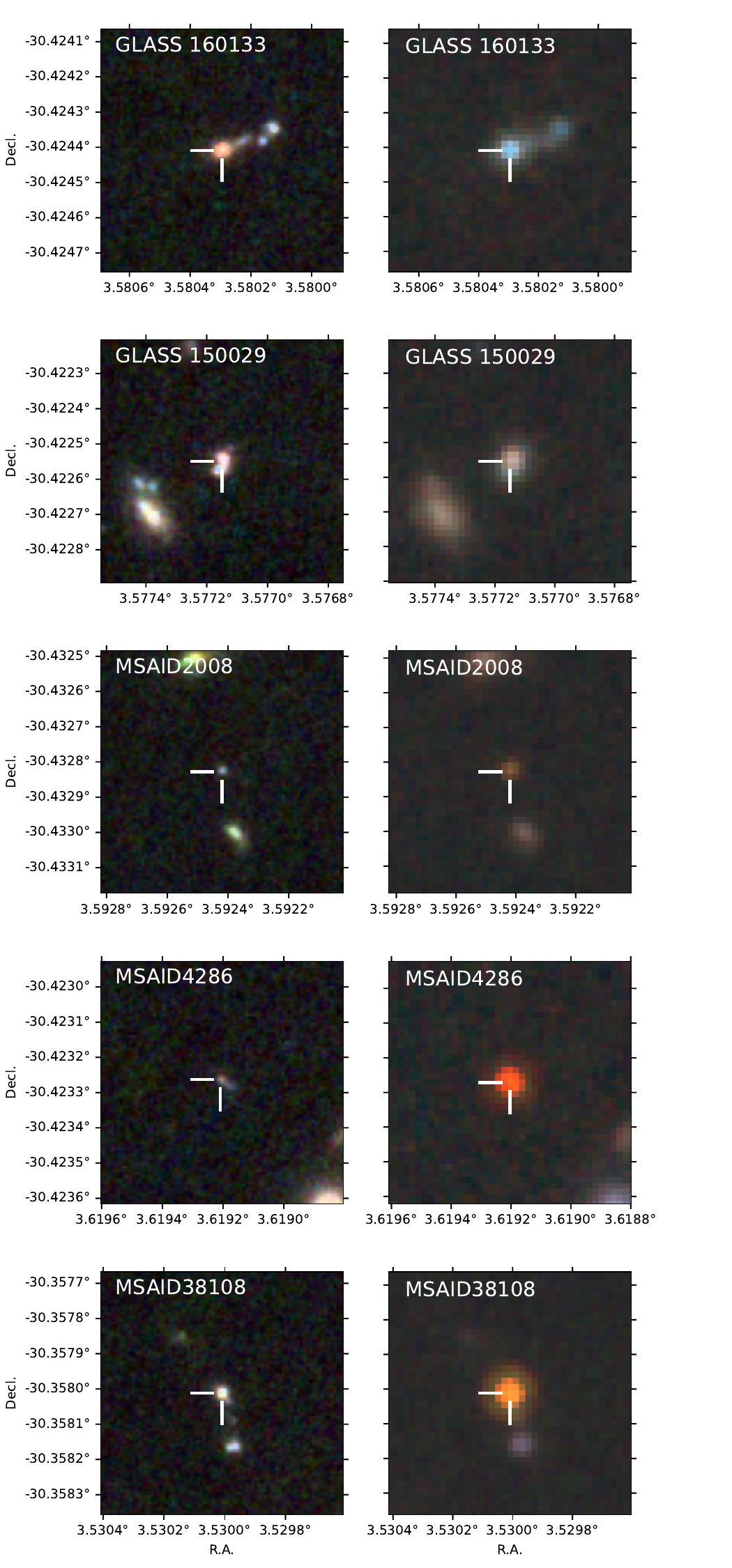}
}
 \caption{$2''.52 \times 2''.52$ NIRCam color image stamps of the {\it JWST} broad H$\alpha$ emitters studied in this work, created from the all-epoch coadd images (see Section~\ref{sec:nircam_image_processing}). The color coding is based on the F115W (blue), F150W (green), and F200W (red) images in the left column, and F277W, F356W, and F444W images in the right column. 
 The marked positions indicate the putative AGN locations (brightest spots in each of the objects).}
 \label{fig:colorimage}
\end{figure}

\subsection{Sample selection}
\label{sec:sample_selection}

\begin{deluxetable*}{ccccccccccc}
\tablecolumns{11}
\tablecaption{Properties of the {\it JWST} broad H$\alpha$ emitters studied in this work. \label{tbl:targets}}
\tablehead{
  \colhead{Name} & 
  \colhead{LRD?} & 
  \colhead{$\text{ID}_{\text{UNCOVER}}$} & 
  \colhead{R.A.}& 
  \colhead{Decl.}&
  \colhead{$z_{\text{spec}}$} &
  \colhead{$\mu$} &
  \colhead{$L_{\text{H}\alpha, \text{broad}}$} &
  \colhead{$\text{FWHM}_{\text{H}\alpha, \text{broad}}$} & 
  \colhead{$M_{\text{BH}}$} & 
  \colhead{Reference} \\
  \colhead{} & 
  \colhead{} & 
  \colhead{} & 
  \colhead{} & 
  \colhead{} & 
  \colhead{} & 
  \colhead{} &
  \colhead{$(10^{42}~\text{erg}~\text{s}^{-1})$} &
  \colhead{$(\text{km}~\text{s}^{-1})$} &
  \colhead{$(10^{7}~M_{\odot})$} &
  \colhead{}
  }
\startdata
  GLASS~160133 & N & 13322 & 00:14:19.271 & -30:25:27.87 & 4.015 & 1.680 & $0.67^{+0.06}_{-0.06}$ & $1028^{+19}_{-13}$ & 0.17 & (1) \\
  GLASS~150029 & N & 14033 & 00:14:18.515 & -30:25:21.17 & 4.583 & 1.676 & $0.47^{+0.04}_{-0.03}$ & $1429^{+10}_{-67}$ & 0.27 & (1) \\
  MSAID2008    & Y & 10065 & 00:14:22.181 & -30:25:58.18 & 6.740 & 1.691 & $2.5^{+0.5}_{-0.5}$ & $1200^{+430}_{-430}$ & 0.48 & (2)\\ 
  MSAID4286    & Y & 13742 & 00:14:28.609 & -30:25:23.78 & 5.840 & 1.615 & $23^{+1}_{-1}$ & $2900^{+1040}_{-1040}$ & 10 & (2)\\
  MSAID38108   & Y & 50494 & 00:14:07.202 & -30:21:28.84 & 4.960 & 1.588 & $26^{+1}_{-1}$ & $4100^{+1980}_{-1980}$ & 22 & (2)\\
\enddata
\tablenotetext{}{The `LRD?' column indicates whether each broad H$\alpha$ emitter is classified as an LRD (Y) or a non-LRD (N). The UNCOVER ID, sky coordinates (ICRS), and gravitational lensing magnification factor $\mu$ are taken from the DR3 photometric catalog \citep{sue24}. The astrometry of the UNCOVER photometric catalog is based on the F444W-band image calibrated against the GAIA DR3. The spectroscopic redshifts $z_{\text{spec}}$, broad H$\alpha$ luminosity $L_{\text{H}\alpha, \text{broad}}$, and broad H$\alpha$ line widths $\text{FWHM}_{\text{H}\alpha, \text{broad}}$ are based on the values given in the references: (1) \cite{har23}; (2) \cite{gre24}. $L_{\text{H}\alpha, \text{broad}}$ is dereddened and demagnified, and $M_{\text{BH}}$ is calculated using the \cite{gre05} relation (see Section~\ref{sec:sample_selection} for the calculation). $\text{FWHM}_{\text{H}\alpha, \text{broad}}$ is the line velocity width deconvolved with the instrumental line spread function.}
\end{deluxetable*}

\begin{figure*}[tbp]
\center{
\includegraphics[clip, width=6.8in]{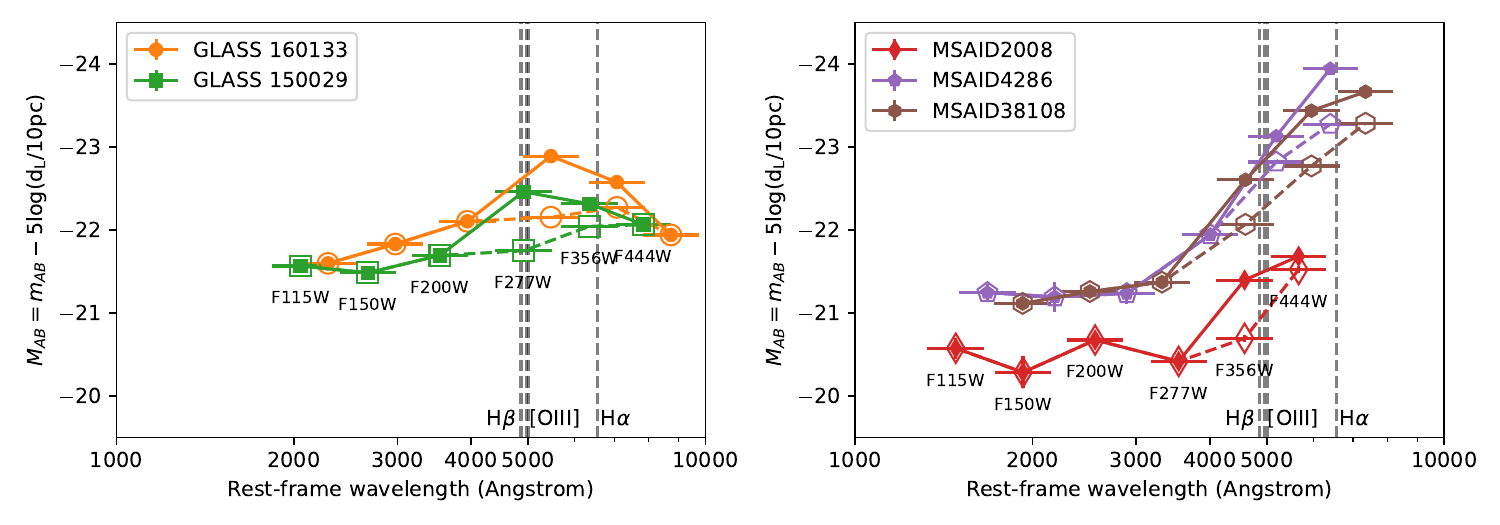}
\includegraphics[clip, width=6.8in]{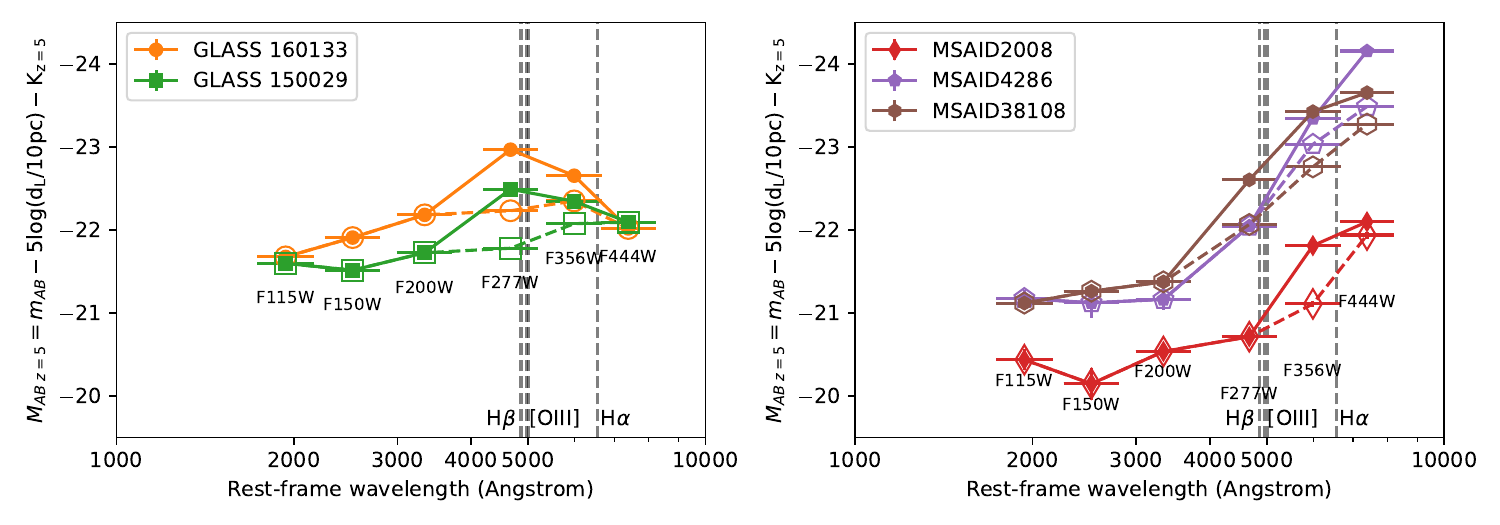}
}
 \caption{Top left: the rest-frame UV-optical SEDs of non-LRD broad H$\alpha$ emitters GLASS~160133 and GLASS~150029 based on the {\it JWST}/NIRCam wide-band photometry from the UNCOVER DR3 ``SUPER'' photometry catalog. 
 Top right: same as the left panel, but for LRD broad H$\alpha$ emitters MSAID2008, MSAID4286, and MSAID38108.
 The vertical axis is the observed AB magnitudes shifted by $5\log(d_{L}/10~\text{pc})$ where $d_{L}$ is the luminosity distance, and the horizontal axis is $\lambda_{\text{rest}} = \lambda_{\text{obs}}/(1+z_{\text{spec}})$. 
 The filled symbols denote the observed magnitudes, and open symbols denote the magnitudes after subtracting the H$\beta$, [\ion{O}{3}], and H$\alpha$ emission line flux contribution (Table~\ref{tbl:line_contamination}).
 Bottom panels: same as the top panels, but the absolute magnitudes are $K$-corrected relative to a reference redshift of $z=5$ (see text in Section 2.1).}
 \label{fig:sed}
\end{figure*}

In this work, we examine the rest-frame UV-optical flux variability of the broad H$\alpha$ emitters discovered by the NIRSpec/Micro-Shutter Assembly (MSA) spectroscopic surveys presented by \cite{har23} and \cite{gre24} in the field of the Hubble Frontier Field (HFF) strong lensing cluster Abell~2744, which is the target field of the Grism Lens Amplified Survey from Space \citep[GLASS;][]{tre22,par23} and the Ultradeep NIRSpec and NIRCam ObserVations before the Epoch of Reionization \citep[UNCOVER;][]{bez22}\footnote{\href{https://jwst-uncover.github.io/}{https://jwst-uncover.github.io/}}.
To constrain the stochastic variability of AGNs, it is essential to combine multi-band, multi-epoch photometric measurements (Section~\ref{sec:likelihood}).
Therefore, the targets for the variability search were chosen based on the availability of multi-epoch, multi-band {\it JWST}/NIRCam wide-band (F115W, F150W, F200W, F277W, F356W, and F444W) imaging data in the Mikulski Archive for Space Telescopes (MAST) \footnote{\href{https://mast.stsci.edu/search/ui/\#/jwst}{https://mast.stsci.edu/search/ui/\#/jwst}} at the time of this writing.

We found that five broad H$\alpha$ emitters (GLASS~160133 and GLASS~150029 from \citealt{har23}; and MSAID2008, MSAID4286, and MSAID38108 from \citealt{gre24}) meet the aforementioned multi-band multi-epoch conditions (see Figure~\ref{fig:abell2744}, Tables~\ref{tbl:summary_data} and ~\ref{tbl:targets}, and Appendix~\ref{sec:log_of_observation}).
Figure~\ref{fig:colorcolorspace} shows the NIRCam broad-band color–color
diagrams of the five broad H$\alpha$ emitters, compared with those of the entire samples of the LRD and non-LRD broad H$\alpha$ emitters presented in Table~2 of \cite{har23} and Table~2 of \cite{gre24}.
The broad-band photometry is taken from the ``SUPER'' photometric catalogs for CEERS \citep[Public Release 1.1:][]{wri24} and UNCOVER \citep[DR3:][]{wea24,sue24}\footnote{UNCOVER\_DR3\_LW\_SUPER\_catalog.fits v~5.2.0 is available at \href{https://jwst-uncover.github.io/DR3.html}{https://jwst-uncover.github.io/DR3.html}, and CEERS\_PR1.1\_LW\_SUPER\_CATALOG.fits v~2.0.1 is available at \href{https://zenodo.org/records/11658282}{https://zenodo.org/records/11658282}. CEERS~01244, CEERS~01665, and CEERS~00717 in \citealt{har23} are outside the CEERS multi-band imaging footprint, and therefore, their broad-band photometry is unavailable in the CEERS photometry catalog.}, in which the photometry is performed using circular apertures with optimally assigned aperture sizes \citep[see Section 4.3 of][]{wea24}.
The photometric LRD selection criteria defined by \cite{gre24} are also shown.
As seen in Figure~\ref{fig:colorcolorspace}, GLASS~160133 and GLASS~150029 exhibit flatter long-wavelength SEDs than the LRDs, and are therefore classified as non-LRD broad H$\alpha$ emitters (Section~\ref{sec:sample_nonlrd}).
GLASS~160133 and GLASS~150029 appear to have particularly blue optical slopes compared to the entire \cite{har23} sample; however, as confirmed below, this is caused by strong emission-line contamination in the F277W filter.
In contrast, MSAID2008, MSAID4286, and MSAID38108 display typical LRD colors and are thus classified as LRD broad H$\alpha$ emitters (Section~\ref{sec:sample_lrd}).

We grouped the NIRCam data for these objects into separate observation dates to define `epochs' of observations for each of the objects, as summarized in Table~\ref{tbl:summary_data} and shown in Figure~\ref{fig:abell2744} (see Appendix~\ref{sec:log_of_observation} for details).
Table~\ref{tbl:targets} summarizes the properties of the five {\it JWST} broad H$\alpha$ emitters studied in this work, and Figure~\ref{fig:colorimage} shows their color image stamps.
All of the targets are located at high redshifts ($z>4$) and in the outskirts of the Abell~2744 cluster (Figure~\ref{fig:abell2744}), and the gravitational lensing magnification $\mu$ by the cluster's gravitational potential is estimated to be insignificant \citep[based on the UNCOVER strong lensing model {\tt v1.1};][]{fur23,wea24,sue24}.
Throughout the paper, we do not apply the magnification and reddening corrections unless otherwise stated.

Figure~\ref{fig:sed} shows the rest-frame optical SEDs of the five {\it JWST} broad H$\alpha$ emitters based on the {\it JWST}/NIRCam wide-band photometry from the UNCOVER DR3 ``SUPER'' photometry catalog.
The top panels show the absolute magnitude $M = m - \text{DM}$ as a function of rest-frame wavelength, where $\text{DM} = 5\log(d_{L}/10~\text{pc})$ indicates the distance modulus.
From this figure, it is clear that the F115W$-$F444W bands used in our variability analysis provide broad coverage around the Balmer break at $\lambda_{\text{rest}}=3647$\AA.
For a fair comparison of the brightness of the five {\it JWST} broad H$\alpha$ emitters at different redshifts, we also calculated $K$-corrected absolute magnitudes as $M_{z=5} = m - \text{DM} - K_{z=5}$, where $K_{z=5}$ denotes the $K$-correction relative to $z=5$ \citep[i.e., a correction factor to transfer the observed broad-band flux into the luminosity in the same band at $z=5$; see Appendix~B of][]{ros13}.
Here, we adopted $z=5$ as a representative central redshift of the five {\it JWST} broad H$\alpha$ emitters.
To compute the $K$-correction, we adopted the following double power-law spectral model for the LRDs \citep[Section~4.4 of][]{gre24}:
$f_{\lambda} \propto \lambda^{\beta}$ at $\lambda < 3647\text{\AA}$ and $f_{\lambda} \propto \lambda^{\alpha}$ at $\lambda > 3647\text{\AA}$ in the rest-frame, with spectral slopes fixed to the mean values measured for the LRD broad H$\alpha$ emitters: $\beta = -1.5$ and $\alpha = 0.5$ \citep{gre24}.
For simplicity, we adopted a single power-law model $f_{\lambda} \propto \lambda^{\beta}$ with $\beta = -1.4$ for the non-LRDs broad H$\alpha$ emitters \citep[as adopted by][]{mad24}.
We note that the UV power-law slope of $\beta \simeq -1.5$ is consistent with the typical UV-optical slope observed in broad-line quasars \citep{van01}.
The $K$-corrected SEDs are shown in the bottom panels of Figure~\ref{fig:sed}.
Since the broad-band photometry is affected by the strong emission lines, such as H$\beta$, \ion{O}{3}, and H$\alpha$ \citep{wil24}, we approximately evaluated the line contamination fraction for each object in each of the NIRCam broad-bands by using the spectral fitting results given in the literature \citep{har23,gre24} convolved with the filter transmission curves\footnote{NIRCam transmission curves version 5.0 (November 2022) available at \href{https://jwst-docs.stsci.edu/jwst-near-infrared-camera/nircam-instrumentation/nircam-filters}{https://jwst-docs.stsci.edu/jwst-near-infrared-camera/nircam-instrumentation/nircam-filters}.} (Table~\ref{tbl:line_contamination}).
The line-subtracted continuum SEDs are also shown in Figure~\ref{fig:sed}.

\subsubsection{Non-LRD broad H$\alpha$ emitters GLASS~160133 and GLASS~150029}
\label{sec:sample_nonlrd}

GLASS~160133 and GLASS~150029 were discovered via direct broad H$\alpha$ emitter search in the NIRSpec galaxy sample \citep{har23}.
GLASS~160133 and GLASS~150029 have extended morphologies (Figure~\ref{fig:colorimage}), and their rest-frame optical SEDs are much flatter than LRDs (Figures~\ref{fig:colorcolorspace} and \ref{fig:sed}); thus, they are non-LRD broad H$\alpha$ emitters.
In any case, under the assumption that the broad H$\alpha$ line is the direct AGN BLR emission, the AGN continuum emission should be contributing to the observed SED to a certain extent \citep[at least at $\lambda_{\text{rest}} \gtrsim 6000\text{\AA}$ even when the AGN is reddened; e.g.,][]{wan24,mad24}.
For GLASS~160133 and GLASS~150029, we assume that the putative AGN position (from which the broad H$\alpha$ originates) corresponds to the brightest point-like source in each of the objects as indicated in Figure~\ref{fig:colorimage}.

Assuming the \cite{cal00} dust extinction law with $E(B-V)$ estimated from the narrow-line Balmer decrement, the extinction-corrected demagnified broad H$\alpha$ luminosities of GLASS~160133 and GLASS~150029, recalculated from the values given in \cite{har23}, are $L_{\text{H}\alpha, \text{broad}} = 6.7\times 10^{41}~\text{erg}~\text{s}^{-1}$ and $4.7\times 10^{41}~\text{erg}~\text{s}^{-1}$ with the FWHMs of $1028$~km~s$^{-1}$ and $1429$~km~s$^{-1}$, respectively (Table~\ref{tbl:targets}).
Based on the \cite{gre05} relation between the BH mass and broad H$\alpha$ FWHM and luminosity, the BH masses of GLASS~160133 and GLASS~150029 are estimated as $M_{\text{BH}} = 1.7 \times 10^{6}~M_{\odot}$ and $2.7 \times 10^{6}~M_{\odot}$, respectively.
The H$\alpha$ luminosity and $M_{\text{BH}}$ of these objects indicate that they are accreting at sub-Eddington rates.
The rest-frame equivalent widths (EWs) of the broad H$\alpha$ emission component are 110\AA, 68\AA\ for GLASS 160133 and GLASS 150029, respectively. 
These EWs are comparable to those of typical unobscured AGNs \citep[e.g.,][]{mai24}, suggesting that the host galaxy's stellar light does not significantly contaminate the continuum in the H$\alpha$ spectral region.

\subsubsection{LRD broad H$\alpha$ emitters MSAID2008, MSAID4286, and MSAID38108}
\label{sec:sample_lrd}

MSAID2008, MSAID4286, and MSAID38108 were LRDs selected by the UNCOVER NIRCam color and morphology selection and spectroscopically followed up by NIRSpec \citep{lab23,gre24}.
\cite{gre24} found clear broad H$\alpha$ emission lines in MSAID4286 and MSAID38108, and a hint of a broad H$\alpha$ emission line in MSAID2008 at low significance (Table~\ref{tbl:targets}).
As shown in Figure~\ref{fig:colorimage}, these LRDs are (by definition) observed as compact IR-bright objects.
As a prototypical LRD broad H$\alpha$ emitter, \cite{gre24} show in their Fig.~6 that MSAID4286's broad H$\alpha$ emission line ($\text{FWHM}_{\text{H}\alpha, \text{broad}} \sim 2900$~km~s$^{-1}$) and red optical continuum at $\lambda_{\text{rest}} > 4000\text{\AA}$ are best explained by the moderately-obscured ($A_{V}=2.7$~\text{mag}) AGN model, whereas the rest-frame UV light is dominated by either the host galaxy's stellar light or dust/electron-scattered AGN accretion disk emission.
Also, \cite{lab23} suggest that, given the ALMA 1.2~mm non-detections, the rest-frame UV-optical SEDs of MSAID2008 and MSAID38108 cannot be explained without invoking a reddened AGN component in combination with a blue stellar or scattered AGN component in the UV wavelengths ($\text{F115W} - \text{F200W} \sim 0$~mag).

However, there are counterarguments to interpreting LRDs within an AGN scenario. 
In particular, the MIR non-detections by {\it JWST}/MIRI indicate that their optical SEDs cannot be easily reproduced by a simple dust-reddened AGN model.
If the $v$-shaped spectra observed in some LRDs are interpreted as Balmer breaks, they further challenge the obscured AGN scenario proposed by \cite{lab23} \citep[see also][]{wan24}.
In this paper, we do not evaluate specific spectral models; instead, we explore the variability while allowing for the broader possibility that these LRDs may host ``red AGNs'' whose SEDs differ from those of known type $1-1.9$~AGNs rather than being ordinary AGNs simply reddened by dust.
We remain agnostic about the relative flux contributions of the host galaxy and the AGN, and we will demonstrate that, as long as the host galaxy contribution lies within a reasonable range, the variability constraints provide a powerful means of assessing the validity of the AGN scenario (see Section~\ref{sec:comparison_with_know_AGNs}).

From the broad H$\alpha$ line fluxes given in Table~3 of \cite{gre24}, the extinction-corrected demagnified broad H$\alpha$ luminosities are calculated as 
$L_{\text{H}\alpha, \text{broad}} = 2.49 \times 10^{42}~\text{erg}~\text{s}^{-1}$, $2.28 \times 10^{43}~\text{erg}~\text{s}^{-1}$, and $2.56 \times 10^{42}~\text{erg}~\text{s}^{-1}$ ($A_{V} = 2.4$, $2.7$, and $3.3$~mag assuming the SMC extinction curve) for MSAID2008, MSAID4286, and MSAID38108, respectively (Table~\ref{tbl:targets}).
The BH masses of MSAID2008, MSAID4286, and MSAID38108 are estimated from the \cite{gre05} relation as $M_{\text{BH}} = 4.8 \times 10^{6}~M_{\odot}$, $1.0 \times 10^{8}~M_{\odot}$, and $2.2 \times 10^{8}~M_{\odot}$, respectively.
The H$\alpha$ luminosity and $M_{\text{BH}}$ of these objects indicate that they are accreting at $\sim 10-100$\% of the Eddington limit.
The large rest-frame equivalent widths (EWs) of the broad H$\alpha$ emission component (470\AA, 780\AA, and 426\AA\ for MSAID2008, MSAID4286, and MSAID38108, respectively) suggest that the AGN continuum dominates the continuum at least in the H$\alpha$ spectral region \citep[see, e.g.,][]{mai24}.

\begin{deluxetable}{ccc}
\tablecolumns{3}
\tablecaption{Correction factors for the emission line contamination in the NIRCam wide-band filters \label{tbl:line_contamination}}
\tablehead{
  \colhead{Name} & 
  \colhead{Band} & 
  \colhead{$\Delta m$}\\
  \colhead{} & 
  \colhead{} & 
  \colhead{(mag)}
  }
\startdata
 GLASS 160133 & F277W & 0.7359 \\
 GLASS 160133 & F356W & 0.3037 \\
 GLASS 160133 & F444W & 0.0000 \\ \hline
 GLASS 150029 & F277W & 0.7110 \\
 GLASS 150029 & F356W & 0.2699 \\
 GLASS 150029 & F444W & 0.0000 \\ \hline
 MSAID2008    & F277W & 0.0000 \\
 MSAID2008    & F356W & 0.7015 \\
 MSAID2008    & F444W & 0.1592 \\ \hline
 MSAID4286    & F277W & 0.0001 \\
 MSAID4286    & F356W & 0.3105 \\
 MSAID4286    & F444W & 0.6684 \\ \hline
 MSAID38108   & F277W & 0.5372 \\
 MSAID38108   & F356W & 0.6657 \\
 MSAID38108   & F444W & 0.3788 \\ \hline
\enddata
\tablenotetext{a}{Magnitude shifts $\Delta m$ are defined as $m_{\text{cont}} - m_{\text{cont+line}}$, where $m_{\text{cont}}$ is the continuum magnitude, and $m_{\text{cont+line}}$ is the total (continuum + emission-line) magnitude. Flux contributions from the H$\beta$, [\ion{O}{3}], H$\alpha$, and [\ion{N}{2}] emission lines are considered based on the narrow-line + broad-line models in Fig.~1 of \cite{har23} and Fig.~3 of \cite{gre24}.}
\end{deluxetable}

\subsection{NIRCam image processing}
\label{sec:nircam_image_processing}

We downloaded Level 2b calibrated single-exposure data ({\tt cal} products) obtained with the six {\it JWST}/NIRCam wide-band filters (F115W, F150W, F200W, F277W, F356W, and F444W) containing GLASS 160133, GLASS 150029, MSAID2008, MSAID4286, or MSAID38108 from MAST (Table~\ref{tbl:summary_data} and \ref{tbl:observations})\footnote{Accessed on May 9, 2024}.
Single-epoch mosaic images were generated independently for each object as described below (except for GLASS 160133 and GLASS 150029, which appear on the same single-epoch mosaic images).
We define the exposure mid-points in the Modified Julian Date (MJD) of the single-epoch mosaic images (given in Table~\ref{tbl:observations} in Appendix~\ref{sec:log_of_observation}) as the epochs of the observations.
The observer-frame temporal sampling interval of the NIRCam data ranges from $400$ to $500$~days.

First, for the dataset of each filter, the relative world coordinate system (WCS) of the images was aligned by utilizing a function in the {\tt tweakreg} step ({\tt expand\_refcat=True}) in the Stage 3 {\it JWST} image processing pipeline ({\tt calwebb\_image3}, v.1.14.0)\footnote{The {\it JWST} Calibration References Data System (CRDS) context file {\tt jwst\_1230.pmap} was retrieved from the CRDS server.}.
The absolute WCS is aligned to that of the UNCOVER DR3 photometric catalog \citep{sue24} by using the {\tt abs\_refcat} option in the {\tt tweakreg} step.
When the image registration appeared imperfect with the resulting WCS solution, the {\tt tweakreg} step was applied iteratively, using only the objects around the target broad H$\alpha$ emitters (within $<30''$) detected by the {\tt make\_tweakreg\_catalog} process.
Then, the background matching and outlier detection were performed by applying the {\tt skymatch} and {\tt outlier\_detection} steps in the pipeline.
We further corrected the $1/f$ noise \citep{sch20} in each calibrated image by running Chris~Willott's {\tt image1overf.py}\footnote{See \href{https://jwst-docs.stsci.edu/known-issues-with-jwst-data/nircam-known-issues/nircam-1-f-noise-removal-methods}{https://jwst-docs.stsci.edu/known-issues-with-jwst-data/nircam-known-issues/nircam-1-f-noise-removal-methods} and \href{https://github.com/chriswillott/jwst}{https://github.com/chriswillott/jwst}.}.

The calibrated images were then {\it drizzle}-combined to produce mosaic images per filter per epoch by applying the {\tt resample} step in the pipeline.
The final output mosaic images of the {\tt resample} step were configured to have pixel scales of $0.031''$~pixel${}^{-1}$ for short channels (F115W, F150W, and F200W) and $0.063''$~pixel${}^{-1}$ for long channels (F277W, F356W, and F444W).
All the single-epoch mosaic images were projected onto the same WCS with {\tt TAN} projection (north is up, and east is to the left) to achieve proper image registration.

The 2-dimensional global background of the mosaic images was calculated using {\tt photutils.background.MedianBackground} \citep{larry_bradley_2023_7946442} as implemented in the {\tt JWSTBackground} class in the pipeline.
The background box size was set to a small value ($\simeq 1''$) after testing several values to ensure that no spurious local patterns were introduced.
Then, the pixel values of the mosaic images were rescaled to have the flux zero-point of 28~ABmag by using the NIRCam's flux calibration information provided by the pipeline and stored in the fits headers\footnote{\href{https://jwst-docs.stsci.edu/jwst-near-infrared-camera/nircam-performance/nircam-absolute-flux-calibration-and-zeropoints}{https://jwst-docs.stsci.edu/jwst-near-infrared-camera/nircam-performance/nircam-absolute-flux-calibration-and-zeropoints}}.
It turned out that, even after applying the pipeline-based zero-point shift, the absolute photometric calibrations of the multiple mosaic images for each filter were inconsistent with each other by a factor of up to a few percent \citep[see, e.g.,][]{ma24}.
To correct this zero-point shift, the epoch $i$ ($i \geq 2$) mosaic image was multiplied by a correction factor that was measured as a median of the ratios of the overlapping image pixel values of bright stationary objects that are detected in all epochs within 30'' of the target.
The pipeline-produced sigma images (containing resampled uncertainty estimates, given as standard deviation) were accordingly scaled.

Since the multi-epoch multi-band observations were performed at different V3 position angles (Figure~\ref{fig:abell2744}), the PSFs of the mosaic images were rotated by some degrees.
We calculated model PSF images for each mosaic image for a subsequent image differentiation analysis as follows.
We used {\tt webbpsf} (v.1.2.1) to simulate the position-dependent PSFs at the positions of the targets for each of the {\tt cal} images.
The source model for the PSF simulation was assumed to be a power-law spectrum $f_\lambda \propto \lambda^{-1}$ that roughly matches the observed spectral shape of our targets.
Then, the PSF models for each of the mosaic images were produced by {\it drizzle}-combining the {\tt cal} images in the same way as described above but with the simulated PSF models embedded at the positions of the targets.

\subsection{{\it Chandra} X-ray data}
\label{sec:chandra_data}

\begin{figure*}[tbp]
\includegraphics[clip, trim=0.0cm 0.0cm 0.8cm 0.0cm, width=3.4in]{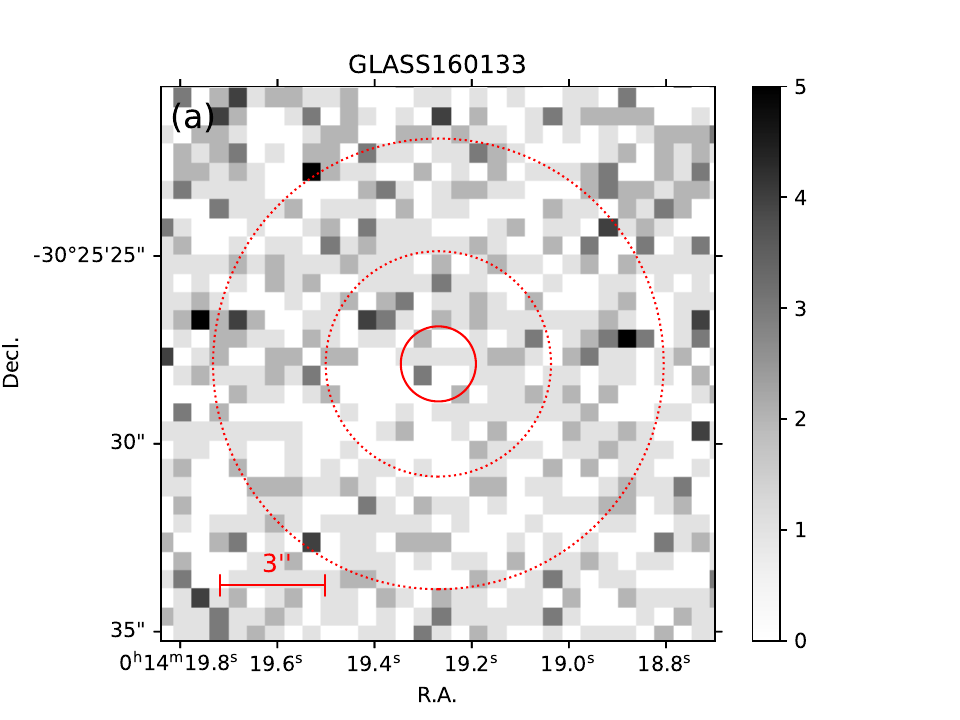}
\includegraphics[clip, trim=0.0cm 0.0cm 0.8cm 0.0cm, width=3.4in]{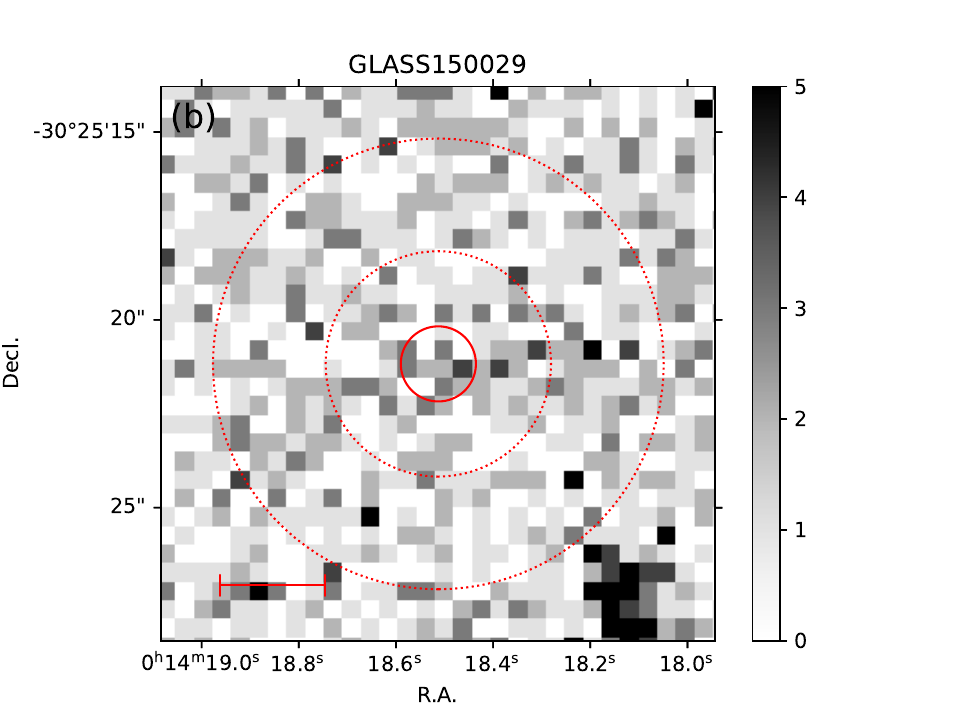}
\includegraphics[clip, trim=0.0cm 0.0cm 0.8cm 0.0cm, width=3.4in]{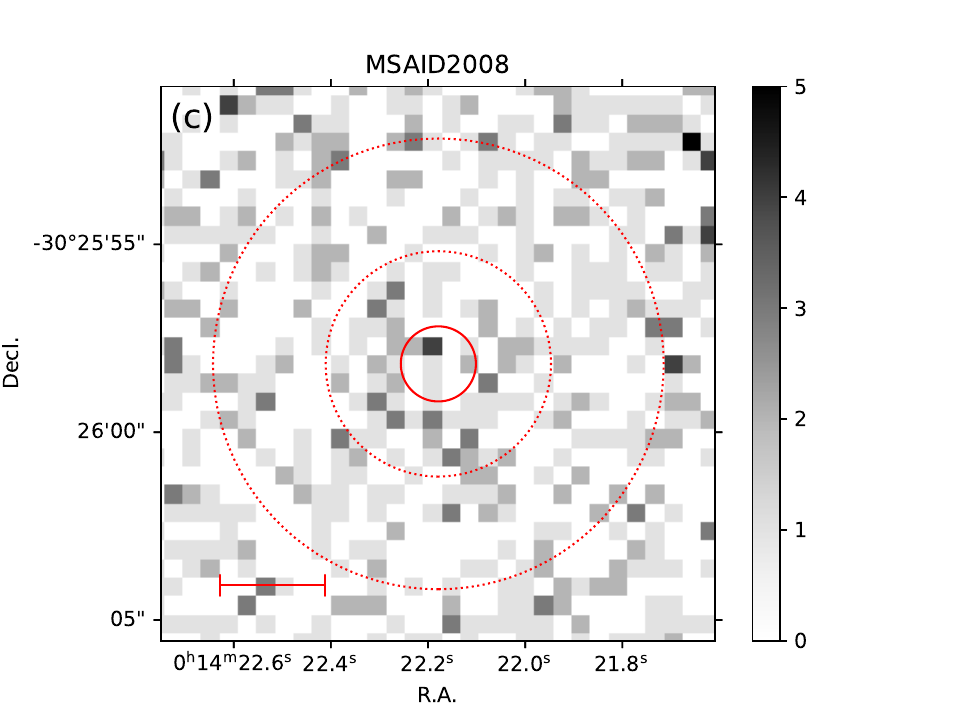}
\includegraphics[clip, trim=0.0cm 0.0cm 0.8cm 0.0cm, width=3.4in]{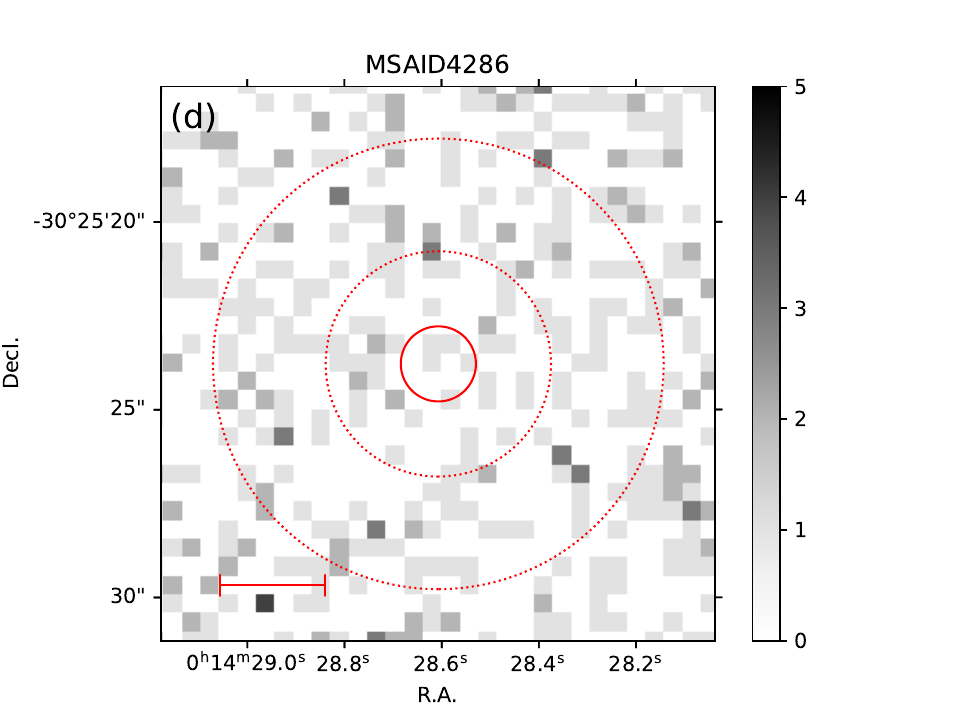}
\includegraphics[clip, trim=0.0cm 0.0cm 0.8cm 0.0cm, width=3.4in]{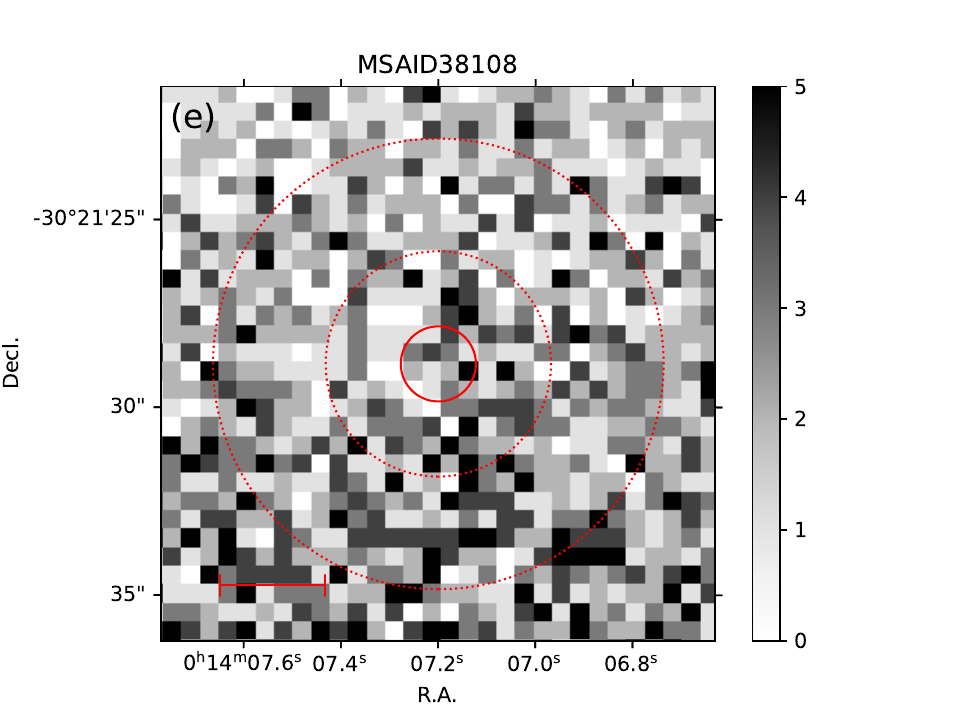}
\includegraphics[clip, trim=-1.2cm 0.0cm 0.8cm 0.0cm, width=3.7in]{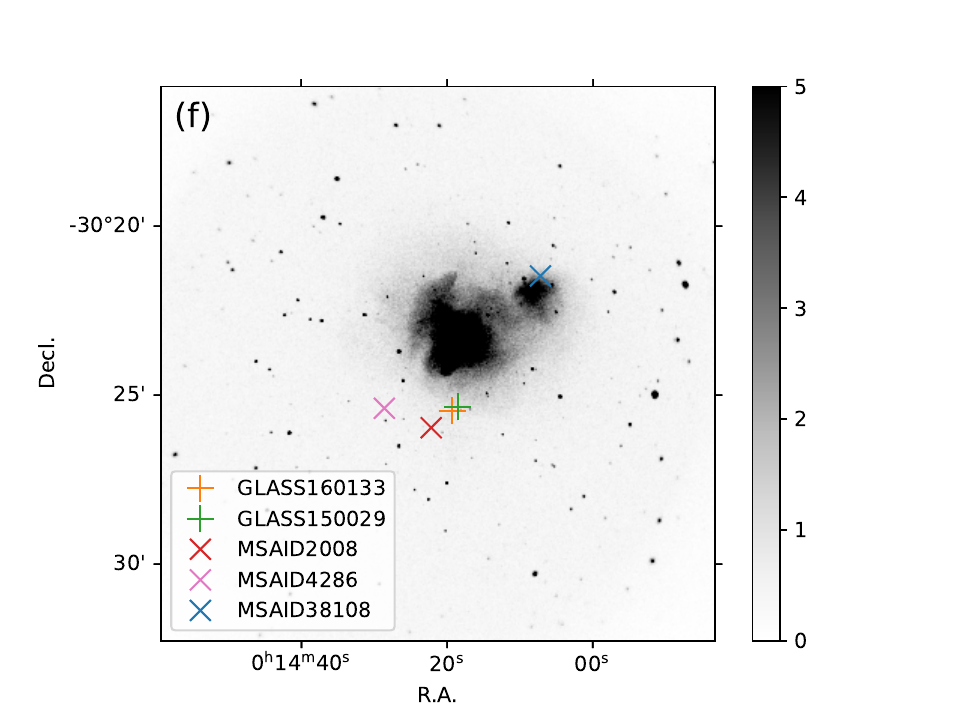}
 \caption{Panels (a)-(e): 2.2~Msec {\it Chandra}/ACIS $0.5-7$~keV X-ray coadded count images centered on the {\it JWST} broad H$\alpha$ emitters ($15'' \times 15''$, $0.492''~\text{pixel}^{-1}$). The colorbar indicates the X-ray counts from 0 (white) to 5 (black) counts. The source extraction aperture ($r=1''$) and background extraction annulus ($r_{\text{in}}=3''$ and $r_{\text{in}}=6''$) are shown as solid and dotted lines, respectively. Panel (f): the X-ray coadded count image over the entire Abell~2744 field. The background level is significantly elevated by the emission of the intracluster medium for objects close to the cluster cores (especially around MSAID38108).}
 \label{fig:chandra_image}
\end{figure*}

\begin{deluxetable}{ccc}
\tablecolumns{3}
\tablecaption{90\% confidence upper limits of the demagnified rest-frame $2-10$~keV X-ray luminosity under the assumption of the power-law X-ray continuum model ($\Gamma=1.7$ or $1.9$). \label{tbl:chandra_upperlimit}}
\tablehead{
  \colhead{Name} & 
  \colhead{$L_{2-10~\text{keV}}$ $(\Gamma=1.7)$} &
  \colhead{$L_{2-10~\text{keV}}$ $(\Gamma=1.9)$}\\
  \colhead{} & 
  \colhead{($10^{42}~\text{erg}~\text{s}^{-1}$)} & 
  \colhead{($10^{42}~\text{erg}~\text{s}^{-1}$)}
  }
\startdata
 GLASS 160133 & $<2.97$ & $<3.52$ \\
 GLASS 150029 & $<4.34$ & $<5.26$ \\
 MSAID2008    & $<6.58$ & $<8.40$ \\
 MSAID4286    & $<5.89$ & $<7.42$ \\
 MSAID38108   & $<15.3$ & $<18.5$ \\
\enddata
\end{deluxetable}

The Abell~2744 field has been observed by {\it Chandra}/Advanced CCD Imaging Spectrometer (ACIS) \citep{wei02} multiple times from 2001 to 2024, especially in the Cycle~23 Very Large Project \#23700107 \citep[e.g.,][]{gou23,bog24,kov24,ana24,cha24}.
The total exposure time amounts to about 2.2~Msec (up to ObsID=29427 obtained on 2024 May 29).
The energy band probed by {\it Chandra}/ACIS ($0.5~\text{keV}-7~\text{keV}$) corresponds to a rest-frame hard X-ray band of $\sim 3~\text{keV}-42~\text{keV}$ for objects at $z \sim 5$.
Thus, the {\it Chandra} observations are a powerful means to explore AGN X-rays unless the observed object is Compton-thick \citep[e.g.,][]{bog24}.
We examined whether the {\it JWST} broad H$\alpha$ emitters studied in this work (Table~\ref{tbl:targets}) were X-ray-detected by using the publicly available {\it Chandra}/ACIS data of Abell~2744\footnote{MSAID2008, MSAID4286, and MSAID38108 correspond to ID=571, 1967, and 28343 in \cite{ana24}, with reported $2-10$~keV X-ray luminosity upper limits of  $\log L_{2-10~\text{keV}} = 43.68, 43.39$, and $43.71$, respectively. Here, we have rederived the X-ray luminosity upper limits for these objects, as well as for GLASS~160133 and GLASS~150029, combining the new {\it Chandra} data in 2024 with the data used by \cite{ana24}.}.

We used {\tt CIAO/download\_chandra\_obsid} to download all the archival ACIS primary and secondary data targetting the Abell~2744 field, and {\tt chandra\_repro} to reprocess them to create level~2 data products with CALDB~v4.11.2.
First, for a visual inspection, a broad-band ($0.5~\text{keV}-7~\text{keV}$) mosaic image was created by combining the reprocessed ACIS data by using {\tt merge\_obs} without binning, using ObsID=8477 as the
reference coordinate system.
Figure~\ref{fig:chandra_image} shows the $0.5-7$~keV coadded clipped count image centered on each of the broad H$\alpha$ emitters.
We did not find any hint of X-ray flux at the positions of the broad H$\alpha$ emitters.

Then, for each object, we extracted the count rate spectra from the individual exposures and combined them into a single coadd spectrum to quantitatively assess the X-ray flux upper limit.
The event files, Ancillary Response Files (ARFs), and Redistribution Matrix Files (RMFs) reprocessed with {\tt chandra\_repro} were input to {\tt specextract} to extract the count rate spectra.
Following \cite{bog24}, a circular aperture of $1$'' in radius was used for the source extraction, and an annular aperture with $3$'' inner and $6$'' outer radii was adopted for the background estimation.
Note that the emission from the intracluster medium in the merging cluster significantly elevates the X-ray background levels for objects close to the cluster cores (the right bottom panel of Figure~\ref{fig:chandra_image}).
The UNCOVER DR3 sky coordinates of the broad H$\alpha$ emitters (Table~\ref{tbl:targets}) were used as the central coordinates of the broad H$\alpha$ emitters. 
Aperture corrections were applied for the ARFs. 
The spectra from the multiple exposures were summed using {\tt combine\_spectra} to create a coadded spectrum for each object.
We obtained the background-subtracted net source count rates in the $0.5-7$~keV band as 
$(-0.6059 \pm 1.467, 3.619 \pm 2.157, 1.666 \pm 1.526, -0.6564 \pm 0.9315, 0.9257 \pm 2.574) \times 10^{-6}$~counts~sec${}^{-1}$ for GLASS~160133, GLASS~150029, MSAID2008, MSAID4286, and MSAID38108, respectively; thus, we conclude that there is no statistically significant X-ray flux in these objects.

Upper limits on the count rates were evaluated from the background counts by using {\tt aplimits}\footnote{\href{https://cxc.cfa.harvard.edu/ciao/threads/upperlimit/}{https://cxc.cfa.harvard.edu/ciao/threads/upperlimit/}} in which the maximum probability of false detection (false positive or Type I error) was set to 0.1, which corresponds to 90\% confidence upper limits \citep[e.g.,][]{kas10}: 
$(2.424, 2.559, 1.970, 1.920, 3.501) \times 10^{-6}$~counts~sec${}^{-1}$ for GLASS~160133, GLASS~150029, MSAID2008, MSAID4286, and MSAID38108, respectively.
The $K$-corrected, unabsorbed rest-frame $2-10$~keV X-ray luminosity upper limits were evaluated with {\tt modelflux} assuming a power-law model with a photon index $\Gamma=1.7$ or $\Gamma=1.9$.
The Galactic hydrogen column density toward the targets \citep[$N_{\text{H}, \text{MW}} \simeq 2 \times 10^{20}~\text{cm}^{-2}$;][]{dic90} was taken from the fits header of each of the coadded spectra, and no obscuration in the target galaxy was assumed.
The calculated upper limits of the $\mu$-demagnified $2-10$~keV X-ray luminosity $L_{2-10~\text{keV}}$ are listed in Table~\ref{tbl:chandra_upperlimit}.

\section{Variability analyses}
\label{sec:analyses}

\subsection{Variability search}

\begin{figure*}[tbp]
\center{
\includegraphics[clip, trim=0.8cm 1.0cm 1.0cm 0.0cm, width=7.2in]{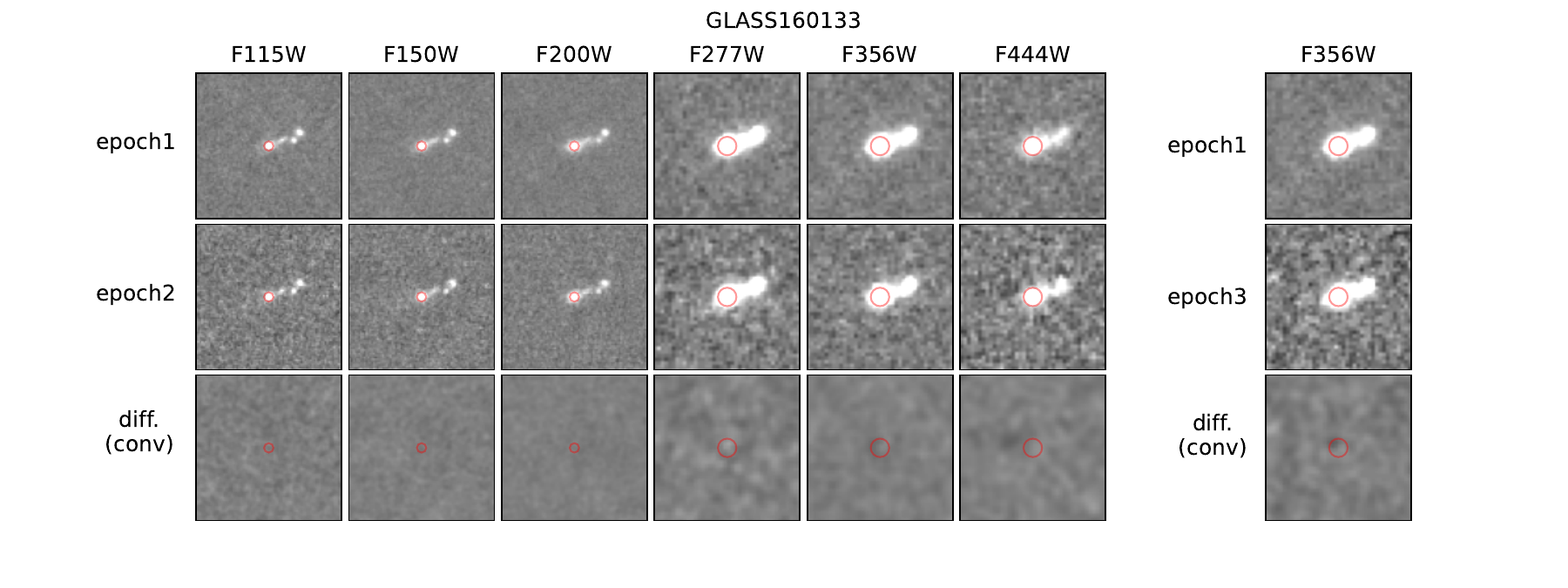}
}
 \caption{Cutouts of the NIRCam single-epoch images of the non-LRD broad H$\alpha$ emitter GLASS~160133. The first and second rows of each panel are the unconvolved epoch 1 and epoch 2/epoch 3 images ($M_{\text{epoch~1}}$, $M_{\text{epoch~2}}$, and $M_{\text{epoch~3}}$), and the third row corresponds to the cross-convolved difference images ($D^{*}$ in Equation~\ref{eqn:cross_convolved_diff}), respectively. The red circles indicate the circular aperture positions ($r=0.08''$ and $0.16''$ for short and long channels, respectively) used for the photometry given in Table~\ref{tbl:aper_photometry}. The gray scale is linear (from $-0.05$ to $0.05$ counts, where 1 count = 28~mag).}
 \label{fig:diff_image}
\end{figure*}

\begin{figure*}[tbp]
\center{
\includegraphics[clip, trim=0.8cm 1.0cm 1.0cm 0.0cm, width=7.2in]{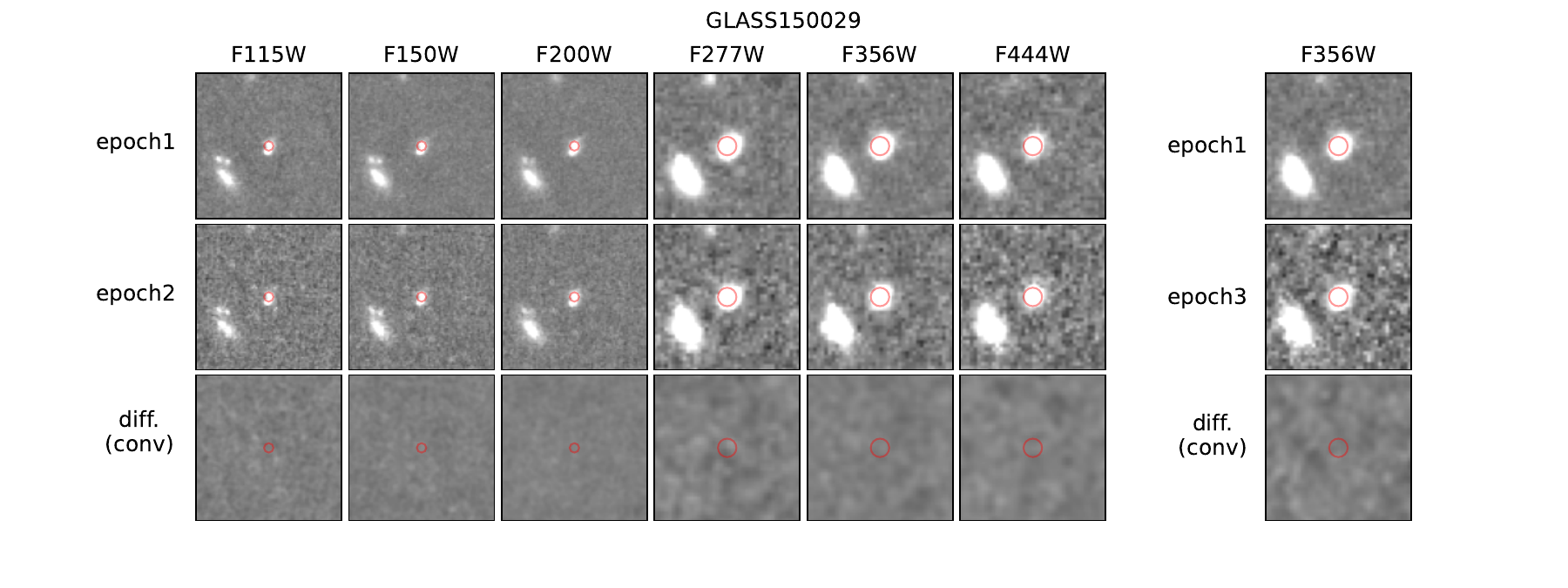}
}
 \caption{Same as Figure~\ref{fig:diff_image}, but for the non-LRD broad H$\alpha$ emitter GLASS~150029.}
 \label{fig:diff_image_2}
\end{figure*}

\begin{figure*}[tbp]
\center{
\includegraphics[clip, trim=0.8cm 1.0cm 1.0cm 0.0cm, width=5.2in]{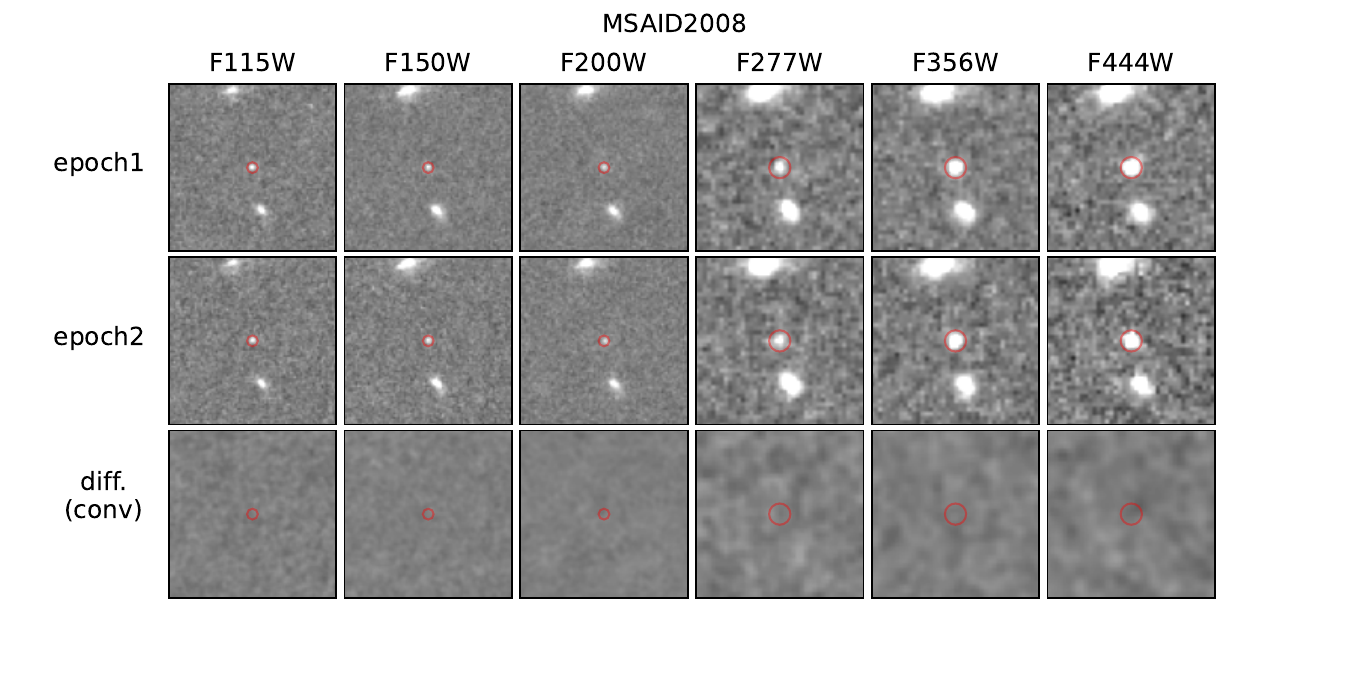}
\includegraphics[clip, trim=0.8cm 1.0cm 1.0cm 0.0cm, width=6.0in]{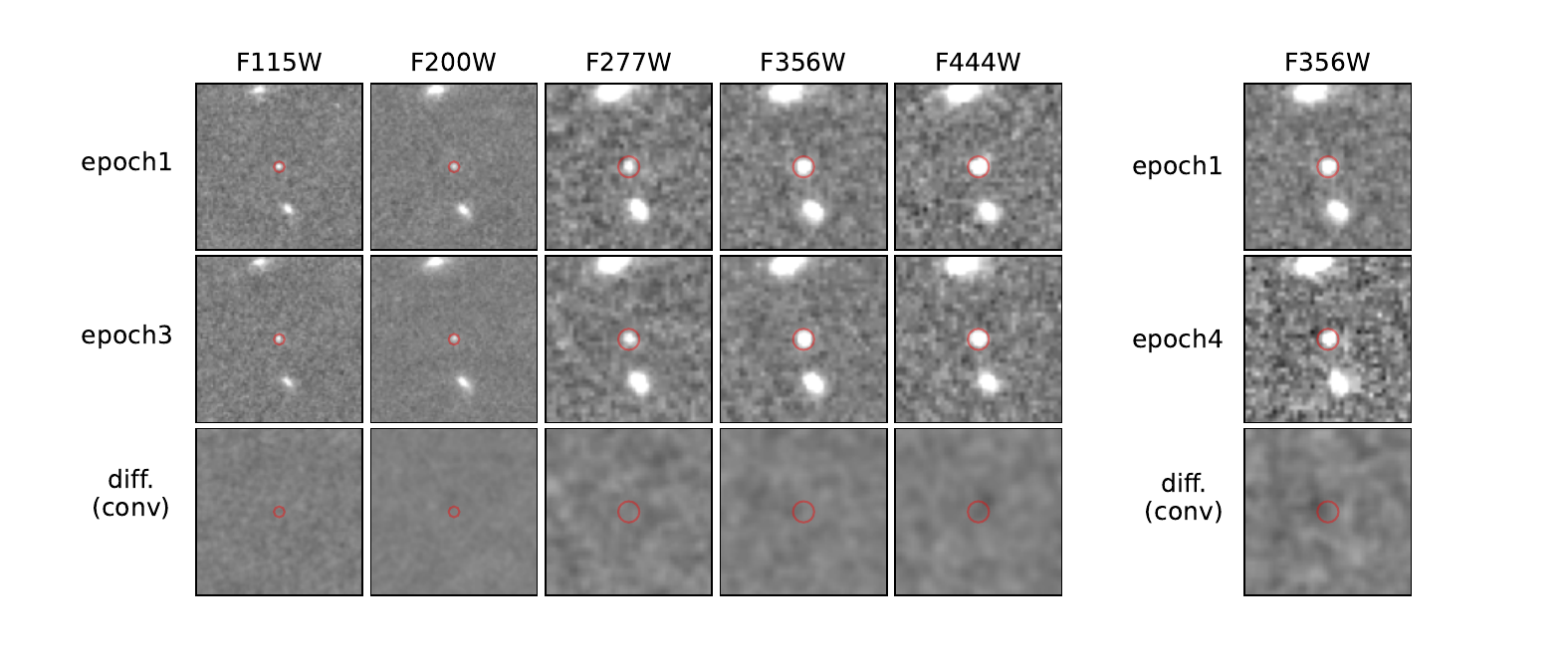}
}
 \caption{Same as Figure~\ref{fig:diff_image}, but for the LRD broad H$\alpha$ emitter MSAID2008. The upper panel shows the epoch~1 and epoch~2 images, and the lower panel shows the epoch~1 and epoch~3/epoch~4 images.}
 \label{fig:diff_image_3}
\end{figure*}

\begin{figure*}[tbp]
\center{
\includegraphics[clip, trim=0.8cm 1.0cm 1.0cm 0.0cm, width=5.2in]{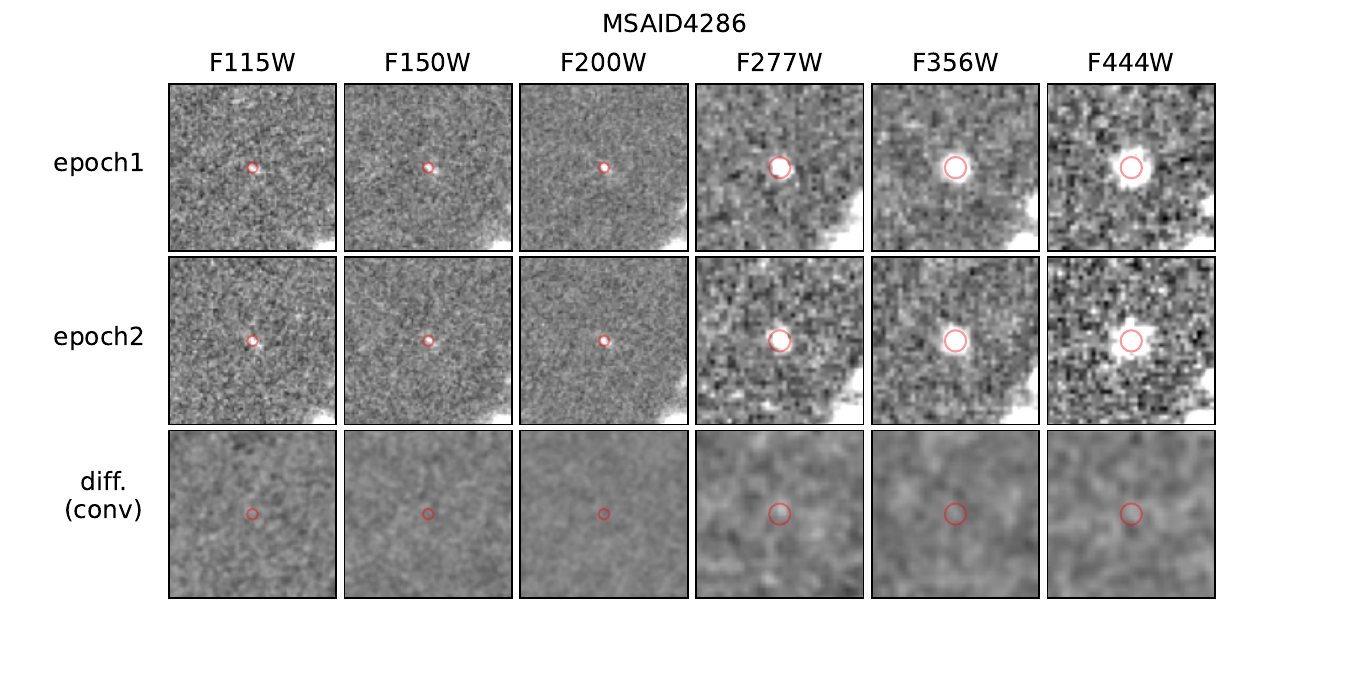}
\includegraphics[clip, trim=0.8cm 1.0cm 1.0cm 0.0cm, width=6.0in]{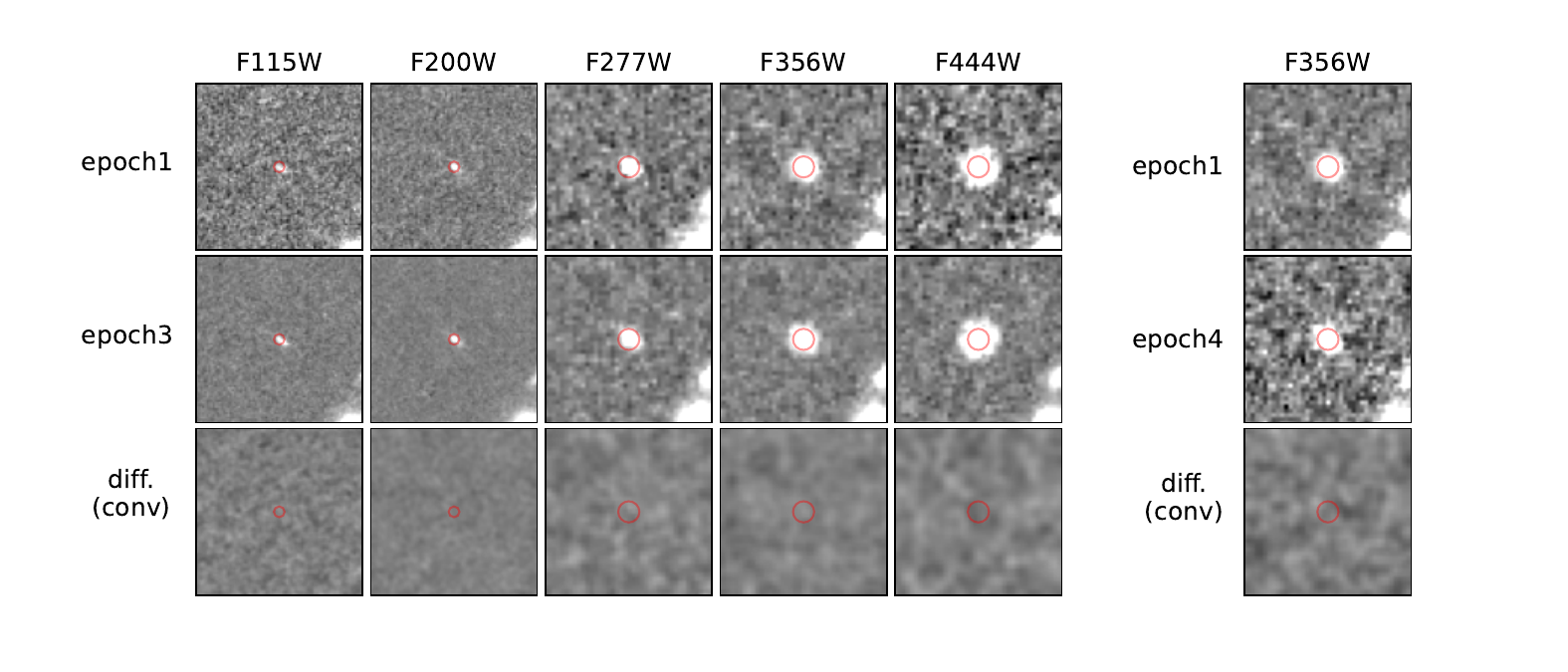}
}
 \caption{Same as Figure~\ref{fig:diff_image_3}, but for the LRD broad H$\alpha$ emitter MSAID4286.}
 \label{fig:diff_image_4}
\end{figure*}

\begin{figure*}[tbp]
\center{
\includegraphics[clip, trim=0.8cm 1.0cm 1.0cm 0.0cm, width=5.2in]{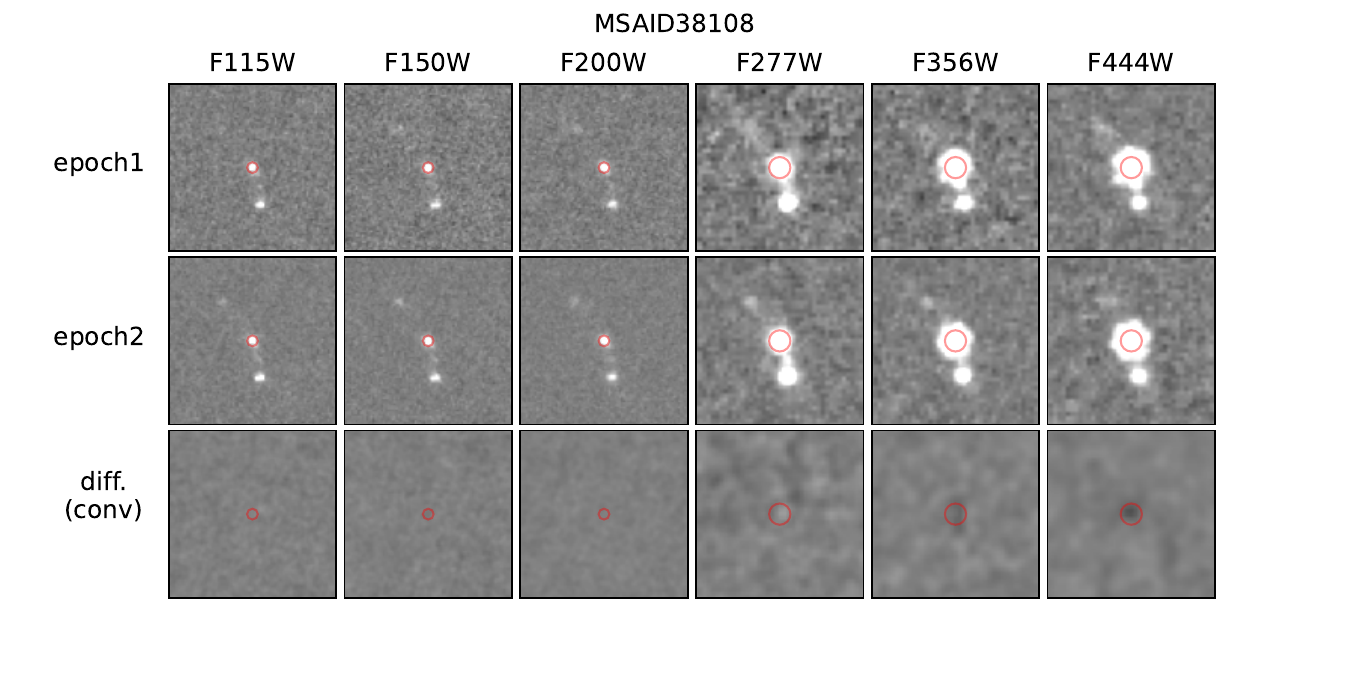}
\includegraphics[clip, trim=0.8cm 1.0cm 1.0cm 0.0cm, width=6.0in]{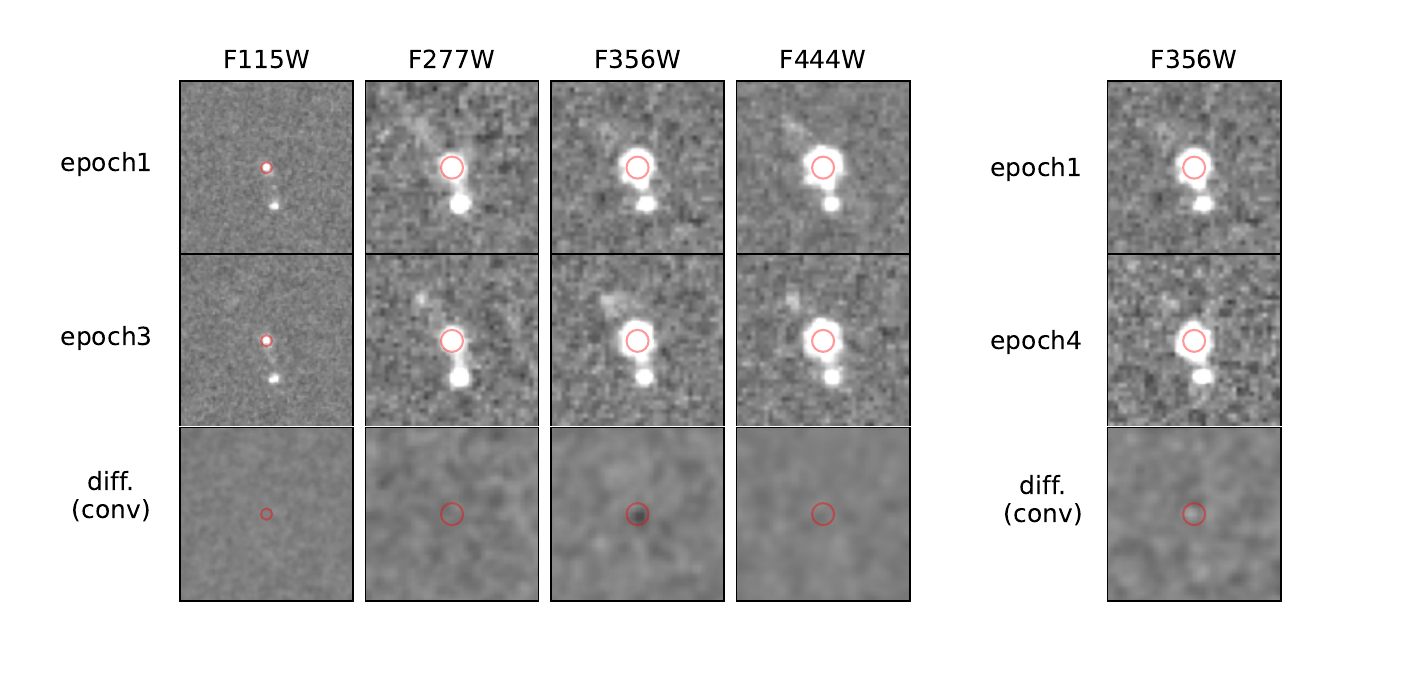}
}
 \caption{Same as Figure~\ref{fig:diff_image_3}, but for the LRD broad H$\alpha$ emitter MSAID38108.}
 \label{fig:diff_image_5}
\end{figure*}

\begin{deluxetable*}{cccccccc}
\tablecolumns{8}
\tablecaption{Multi-epoch aperture AB magnitudes, and significance of the flux variability relative to the epoch~1 flux (Equation~\ref{eqn:sn}). \label{tbl:aper_photometry}}
\tablehead{\colhead{Band} & \colhead{$\text{mag}_{\text{epoch}~1}$} & \colhead{$\text{mag}_{\text{epoch}~2}$} & \colhead{$\text{mag}_{\text{epoch}~3}$} & \colhead{$\text{mag}_{\text{epoch}~4}$} & \colhead{$\text{SN}_{\text{var}, \text{epoch}~2}$} & \colhead{$\text{SN}_{\text{var}, \text{epoch}~3}$} & \colhead{$\text{SN}_{\text{var}, \text{epoch}~4}$}}
\startdata
\multicolumn{8}{c}{ GLASS~160133 } \\
F115W & $26.951\pm0.014$ & $26.933\pm0.028$ &                  &                  & 0.543 &       &          \\
F150W & $26.701\pm0.010$ & $26.697\pm0.021$ &                  &                  & 0.166 &       &          \\
F200W & $26.345\pm0.009$ & $26.339\pm0.015$ &                  &                  & 0.335 &       &          \\
F277W & $25.181\pm0.005$ & $25.178\pm0.007$ &                  &                  & 0.339 &       &          \\
F356W & $25.532\pm0.006$ & $25.549\pm0.009$ & $25.532\pm0.013$ &                  & 1.652 & 0.031 &          \\
F444W & $26.219\pm0.014$ & $26.184\pm0.021$ &                  &                  & 1.327 &       &          \\\hline
\multicolumn{8}{c}{ GLASS~150029 } \\
F115W & $27.381\pm0.021$ & $27.392\pm0.041$ &                  &                  & 0.250 &       &          \\
F150W & $27.431\pm0.019$ & $27.339\pm0.035$ &                  &                  & 2.251 &       &          \\
F200W & $27.179\pm0.016$ & $27.147\pm0.026$ &                  &                  & 0.988 &       &          \\
F277W & $25.807\pm0.008$ & $25.787\pm0.012$ &                  &                  & 1.395 &       &          \\
F356W & $25.955\pm0.008$ & $25.937\pm0.013$ & $25.957\pm0.019$ &                  & 1.198 & 0.112 &          \\
F444W & $26.190\pm0.014$ & $26.183\pm0.022$ &                  &                  & 0.252 &       &          \\\hline
\multicolumn{8}{c}{ MSAID2008 } \\
F115W & $28.684\pm0.089$ & $28.706\pm0.101$ & $28.715\pm0.093$ &                  & 0.160 & 0.231 &          \\
F150W & $28.778\pm0.085$ & $28.651\pm0.102$ &                  &                  & 0.908 &       &          \\
F200W & $28.691\pm0.083$ & $28.782\pm0.097$ & $28.701\pm0.070$ &                  & 0.674 & 0.091 &          \\
F277W & $28.584\pm0.109$ & $28.456\pm0.089$ & $28.466\pm0.082$ &                  & 0.874 & 0.835 &          \\
F356W & $27.676\pm0.048$ & $27.708\pm0.056$ & $27.659\pm0.037$ & $27.761\pm0.102$ & 0.420 & 0.268 & 0.738    \\
F444W & $27.340\pm0.051$ & $27.485\pm0.072$ & $27.496\pm0.055$ &                  & 1.633 & 2.042 &          \\\hline
\multicolumn{8}{c}{ MSAID4286 } \\
F115W & $28.452\pm0.149$ & $28.071\pm0.111$ & $28.301\pm0.066$ &                  & 1.978 & 0.906 &          \\
F150W & $28.297\pm0.118$ & $28.335\pm0.119$ &                  &                  & 0.221 &       &          \\
F200W & $27.934\pm0.074$ & $28.019\pm0.082$ & $27.931\pm0.037$ &                  & 0.744 & 0.048 &          \\
F277W & $26.893\pm0.041$ & $26.732\pm0.042$ & $26.822\pm0.021$ &                  & 2.680 & 1.551 &          \\
F356W & $25.671\pm0.015$ & $25.651\pm0.017$ & $25.663\pm0.008$ & $25.666\pm0.023$ & 0.877 & 0.443 & 0.166    \\
F444W & $24.810\pm0.013$ & $24.778\pm0.013$ & $24.824\pm0.006$ &                  & 1.775 & 0.951 &          \\\hline
\multicolumn{8}{c}{ MSAID38108 } \\
F115W & $27.637\pm0.036$ & $27.595\pm0.021$ & $27.642\pm0.033$ &                  & 1.005 & 0.110 &          \\
F150W & $27.400\pm0.035$ & $27.464\pm0.020$ &                  &                  & 1.490 &       &          \\
F200W & $27.397\pm0.033$ & $27.375\pm0.019$ &                  &                  & 0.570 &       &          \\
F277W & $25.946\pm0.012$ & $25.912\pm0.007$ & $25.951\pm0.012$ &                  & 2.558 & 0.263 &          \\
F356W & $24.975\pm0.006$ & $24.986\pm0.004$ & $24.981\pm0.006$ & $24.961\pm0.008$ & 1.551 & 0.682 & 1.454    \\
F444W & $24.735\pm0.003$ & $24.745\pm0.003$ & $24.730\pm0.004$ &                  & 2.493 & 0.837 &          \\
\enddata
\tablenotetext{a}{See Table~\ref{tbl:summary_data} for the definition of the epochs. Corrections for the extinction and emission line contributions to the aperture magnitudes are not applied. Aperture corrections are applied.}
\end{deluxetable*}

Below, first, we examine the difference images of our targets between the epoch 1 and epoch $i$ ($i \geq 2$) images to demonstrate on the image plane that any flux variations, if present, are too small to be detected by the NIRCam observations.
Since the NIRSpec MSA apertures used for spectroscopy were large enough to contain the multiple components that constitute the complex morphologies of the non-LRD broad H$\alpha$ emitters GLASS~160133 and GLASS~150029 \citep[Figure~\ref{fig:colorimage}; see also Figure~5 of][]{har23}, it should be noted that conservatively speaking we do not know where in those multiple components the broad H$\alpha$ emitters reside.
The image differencing method is especially useful for demonstrating that none of the components around GLASS~160133 and GLASS~150029 exhibit flux variations.
Then, we quantitatively evaluate the significance of the variability at the positions of the brightest spots in each object by performing forced aperture photometry.

\subsubsection{Image differencing}

Consider the image differencing between the epoch 1 and epoch 2 images.
In principle, a proper image subtraction including the pixel areas of extended diffraction features can be achieved by cross-convolving the PSFs of the two images, but accurately reconstructing the PFSs from the mosaic images is exceedingly difficult.
Here, the PSF models simulated with {\tt webbpsf} (Section~\ref{sec:nircam_image_processing}) were used to define the PSF-homogenized cross-convolved mosaic images per filter and per target \citep[e.g.,][]{gal08,hu24}:
\begin{eqnarray}
M_{\text{epoch~1}}^{*} &=& P_{\text{epoch~2}} \otimes M_{\text{epoch~1}}\\
M_{\text{epoch~2}}^{*} &=& P_{\text{epoch~1}} \otimes M_{\text{epoch~2}},
\end{eqnarray}
where $M_{\text{epoch~1}}$ and $M_{\text{epoch~2}}$, respectively, denote the epoch 1 and epoch 2 mosaic images, $P_{\text{epoch~1}}$ and $P_{\text{epoch~2}}$ denote the simulated position (target)-dependent PSF models, and $M_{\text{epoch~1}}^{*}$ and $M_{\text{epoch~2}}^{*}$ denote the cross-convolved mosaic images (epoch 1 and epoch 2, respectively).
The difference image of each filter and target is then defined as the difference between the cross-convolved mosaic images:
\begin{eqnarray}
D^{*} &=& M_{\text{epoch~2}}^{*} - M_{\text{epoch~1}}^{*},
\label{eqn:cross_convolved_diff}
\end{eqnarray}
where its PSF is $P_{\text{epoch~1}} \otimes P_{\text{epoch~2}}$.
The differences between the epoch 1 and epoch 3/epoch 4 images are similarly defined.

Figures~\ref{fig:diff_image}-\ref{fig:diff_image_5} show the cutout images of the epoch~1 ($M_{\text{epoch~1}}$) and of later epochs ($M_{\text{epoch~2}}$, $M_{\text{epoch~3}}$, and $M_{\text{epoch~4}}$), and their cross-convolved difference images ($D^{*}$).
A visual inspection of the difference images reveals no obvious variable point sources in the difference images, suggesting no significant variability in any of the objects in any of the six filters.
It can be confirmed that no excess in the signal significance is present in the difference images at the positions of the targets by evaluating the optimal statistical measure (significance map) for the variability detection \citep[following Equation~25 of][; see their Section~5]{zac16}.
We note that some minor, statistically insignificant subtraction residuals are observed in the difference images, particularly on top of the bright object MSAID38108 imaged at the edge of the NIRCam detectors (Figure~\ref{fig:abell2744}).
Our visual inspection of the unconvolved difference images indicates that these subtraction residuals are caused by imperfections in the PSF modeling and astrometric image registration at subpixel levels. 
However, as quantitatively evaluated below, the total flux summed over circular apertures is neither statistically significantly positive nor negative.
Enhancements in image subtraction, such as addressing undersampled images using the drizzle function and improving astrometric accuracy, will be explored in future investigations.

\subsubsection{Aperture photometry}

Then, we performed aperture photometry on the single-epoch images to evaluate the aperture magnitudes for subsequent analysis of the variability amplitudes.
The aperture photometry was conducted on the unconvolved images ($M_{\text{epoch}~i}$) to avoid pixel correlations caused by the kernel convolution.
We assumed the putative AGN positions as the brightest spots in each of the galaxies (Figure~\ref{fig:colorimage}), on which the aperture photometry was performed.
The aperture radii were fixed to $r=0''.08$ and $r=0''.16$ for the short and long channels, respectively, which were conservatively chosen to be much greater than the NIRCam PSF FWHMs\footnote{\href{https://jwst-docs.stsci.edu/jwst-near-infrared-camera/nircam-performance/nircam-point-spread-functions}{https://jwst-docs.stsci.edu/jwst-near-infrared-camera/nircam-performance/nircam-point-spread-functions}}.
{\tt photutils.aperture} was used to measure the aperture fluxes, utilizing the sigma images for the uncertainty estimation.
Aperture correction factors were calculated by using the PSF models simulated with {\tt webbpsf} (Section~\ref{sec:nircam_image_processing}).

The results of the aperture photometry for the single-epoch images are summarized in Table~\ref{tbl:aper_photometry}.
The $\text{SN}_{\text{var}}$ in Table~\ref{tbl:aper_photometry} is a measure of the significance of the flux variability between epoch~1 and later epochs in each band:
\begin{equation}
\text{SN}_{\text{var}, \text{epoch}~i} = \frac{\left| \text{flux}_{\text{epoch~1}} - \text{flux}_{\text{epoch~}i} \right|}{\sqrt{\sigma_{\text{flux}, \text{epoch~1}}^2+\sigma_{\text{flux}, \text{epoch~}i}^2}},
\label{eqn:sn}
\end{equation}
where $i \geq 2$.
Note that the current error estimates on the magnitudes/fluxes, based on the sigma images, are probably underestimated especially for bright objects due to the omission of additional error sources (e.g., magnitude zero-point and astrometric calibration) that have not been considered.
As a result, the $\text{SN}_{\text{var}}$ is likely overestimated compared to the intrinsic value. Nevertheless, we did not find any significant ($\text{SN}_{\text{var}}>3$) variability in any of the objects. 
We will evaluate the upper limits of the (putative) AGN variability amplitude in Section~\ref{sec:likelihood}, but note that, considering the aforementioned underestimation of the error estimates, these upper limits should be regarded as conservative.

\subsection{Upper limits on the variability amplitude} 
\label{sec:likelihood}

The AGN UV-optical continuum variability is known to follow the Damped Random Walk (DRW) process or Gaussian process with an exponential kernel, which behaves as a random walk on short time scales and asymptotically approaches a finite variability amplitude on long time scales \citep[Appendix~\ref{sec:data_likelihood}; e.g.,][]{kel09,koz10,mac10,zu11,mac12}.
Inter-band time lags of the UV-optical accretion disk continuum are observed to be consistent with light crossing times across the disk radii (at most a few days for local Seyfert galaxies), and on longer time scales multi-band UV-optical light curves exhibit a strong inter-band correlation \citep[e.g.,][]{kok15,fau16,cac18,her20}.
The light curve data likelihood of the DRW model can be expressed by a multivariate Gaussian as given in Equation~\ref{eqn:datalikelihood} \citep[e.g.,][]{ryb92,koz10,zu11}, with which we can perform statistical inference to put constraints on the DRW asymptotic variability amplitude.
We used the multi-band data likelihood of each object to obtain a joint constraint on the asymptotic variability amplitude from the multi-epoch multi-band measurements.
We note that the magnitude shifts due to the extinction corrections, aperture corrections, and emission line contributions do not influence this likelihood analysis (see Appendix~\ref{sec:data_likelihood}).

Following Equation~\ref{eqn:data_vector}, we define the data vector $\vec{y}$ of each object using the aperture photometry magnitudes (Table~\ref{tbl:aper_photometry}).
The rest-frame temporal separations between the two observations are defined as the difference of the MJD-mid (Table~\ref{tbl:observations} in Appendix~\ref{sec:log_of_observation}) scaled by $1/(1+z)$: $\Delta t_{\text{rest}} = \Delta \text{MJD-mid}/(1+z)$ in units of days.
The data likelihood $p(\vec{y}|\vec{\theta})$ given in Equation~\ref{eqn:datalikelihood} defines the likelihood of observing the data $\vec{y}$ of each object given the DRW model parameters $\vec{\theta} = \{\tau_{\text{d}}, \sigma_{\text{d}}(\lambda_{1}), \dots, \sigma_{\text{d}}(\lambda_{N})\}$, where $\lambda_{a}$ is an NIRCam band's pivot wavelength, $\sigma_{\text{d}}(\lambda_{a})$ is the wavelength-dependent asymptotic variability amplitude in units of magnitudes, and $\tau_{\text{d}}$ is the decorrelation time scale in units of days (Equation~\ref{eqn:drw}), and $N = 6$ is the number of bands to be simultaneously fit.
From Bayes' theorem, the posterior distribution of $\vec{\theta}$ given the measurements $\vec{y}$ is defined by multiplying the likelihood $p(\vec{y}|\vec{\theta})$ and a prior distribution of $\vec{\theta}$ denoted as $p(\vec{\theta})$:
\begin{eqnarray}
p(\vec{\theta}|\vec{y}) &\propto& p(\vec{\theta})p(\vec{y}|\vec{\theta}).
\end{eqnarray}
We note that in the literature the asymptotic variability amplitude is more commonly represented by the asymptotic structure function (SF), which is related to each other as (see Appendix~\ref{sec:data_likelihood}):
$\text{SF}_{\infty}(\lambda_{a}) = \sqrt{2}\sigma_{\text{d}}(\lambda_{a})$.

The decorrelation time scale $\tau_{\text{d}}$ cannot be constrained from the current data due to the limited time sampling \citep[e.g.,][]{koz17c}; thus, we assume fixed values of
\begin{equation}
\tau_{\text{d}} = 44.6~\text{days}~\left(\frac{M_{\text{BH}}}{10^7~M_{\odot}}\right)^{0.38}
\end{equation}
following the \cite{bur21}'s empirical relation, using the $M_{\text{BH}}$ estimates given in Section~\ref{sec:sample_selection}.
For GLASS~160133, GLASS~150029, MSAID2008, MSAID4286, and MSAID38108, $\tau_{\text{d}}$ inferred from $M_{\text{BH}}$ given in Table~\ref{tbl:targets} are 23, 27, 34, 107, and 144~days, respectively.
This decorrelation time scale is shorter than or comparable to the rest-frame temporal sampling range of the light curves (Table~\ref{tbl:observations} in Appendix~\ref{sec:log_of_observation}); thus, our light curve data enable us to place constraints on the asymptotic variability amplitude (note that stronger constraints on the asymptotic variability amplitudes are obtained for objects with smaller $M_{\text{BH}}$).

Since no variability was detected, here, we (somewhat arbitrarily) define an upper limit on the variability amplitude for each object as follows.
To obtain multi-band joint constraints on the variability amplitude, a model to relate the variability amplitudes in different bands is needed.
In the literature \citep[e.g.,][]{mac10,bur23}, a power-law function is assumed to model the wavelength dependence of $\sigma_{\text{d}}$:
\begin{eqnarray}
\sigma_{\text{d}}(\lambda) = \sigma_{\text{d}}(4000\text{\AA})\left(\frac{\lambda}{4000\text{\AA}}\right)^{\alpha_{\text{var}}}.
\end{eqnarray}
The reference rest-frame wavelength $4000$~\AA\ is chosen so that $\sigma_{\text{d}}(4000\text{\AA})$ can be compared with literature values of $\text{SF}_{\infty}(4000\text{\AA}) = \sqrt{2}\sigma_{\text{d}}(4000\text{\AA})$ for known AGNs \citep{mac10,sub21,bur23}.
Note that $\sigma_{\text{d}}(\lambda)$ represents the spectral shape of the variable spectral component, which is in general different from the directly observed total spectral flux \citep[e.g.,][]{kok14,rua14,hea23}.
Given the multi-band variability non-detection, $\alpha_{\text{var}}$ is unconstrained from the current data.
A conservative upper limit on $\sigma_{\text{d}}(4000\text{\AA})$ is obtained when $\alpha_{\text{var}} = 0$ is assumed because larger positive/negative $\alpha_{\text{var}}$ requires smaller $\sigma_{\text{d}}(\lambda)$ to be consistent with the multi-band variability non-detection.
Therefore, we fixed $\alpha_{\text{var}} = 0$ to define a conservative upper limit on $\sigma_{\text{d}}(4000\text{\AA})$.

With this parameterization, the multi-band data likelihood can be reinterpreted as a single-parameter function of $\vec{\theta} = \{\log \sigma_{\text{d}}(4000\text{\AA})\}$.
Adopting a uniform prior distribution on $\log \sigma_{\text{d}}(4000\text{\AA})$ over $[-4:0]$, we calculated an upper limit on $\log \sigma_{\text{d}}(4000\text{\AA})$ at the 90\% confidence level from the normalized posterior distribution for each object such that $p(\log \sigma_{\text{d}}(4000\text{\AA}) \leq \log \sigma_{\text{d}}^{*}(4000\text{\AA}) | \vec{y}) = 0.9)$, where $\log \sigma_{\text{d}}^{*}(4000\text{\AA})$ represents the upper limit.
We obtained 
$\log \sigma_{\text{d}}^{*}(4000\text{\AA}) = $
$-2.213$, 
$-1.508$, 
$-1.491$, 
$-1.007$, 
$-1.510$ for 
GLASS~160113, 
GLASS~150029, 
MSAID2008, 
MSAID4286, and 
MSAID38108, respectively, as shown in Figure~\ref{fig:burke_plot} in the form of $\text{SF}_{\infty}^{*}(4000\text{\AA}) = \sqrt{2}\sigma_{\text{d}}^{*}(4000\text{\AA})$.

Using the same likelihood formalism as described above, as we add more data, the upper limits will become progressively tighter. 
This suggests a clear direction for future investigation.

\section{Discussion}
\label{sec:discussion}

\subsection{X-ray weakness of individual objects}
\label{sec:xray_weakness}

\begin{figure}[tbp]
\center{
\includegraphics[clip, trim=0.0cm 0.0cm 0.0cm 0.0cm, width=3.4in]{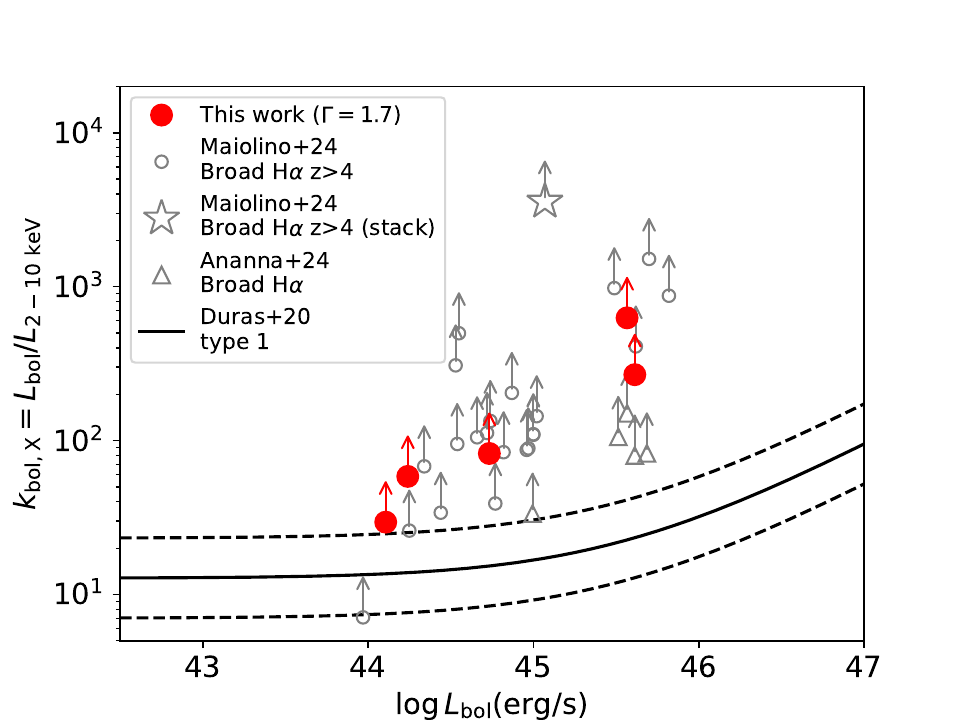}
}
 \caption{$k_{\text{bol}, X}$ as a function of $L_{\text{bol}}$. The data points denote lower limits of $k_{\text{bol}, X}$ because none of them are X-ray detected. Filled circles denote the lower limits for the {\it JWST} broad H$\alpha$ emitters studied in this work (assuming $\Gamma=1.7$ power-law for the X-ray spectral model): GLASS~150029, GLASS~160133, MSAID2008, MSAID4286, and MSAID38108 from left to right. Open circles denote the \cite{mai24}' sample of the LRD and non-LRD broad H$\alpha$ emitters at $z > 4$ discovered by the JADES and GA-NIFS surveys, and the open star symbol denotes the lower limit for the stacked X-ray data. Open triangles denote the \cite{ana24}' UNCOVER sample cross-matched with the \cite{gre24}' sample of the LRD broad H$\alpha$ emitters. The black lines show the best fit and dispersion (1$\sigma$) of the $k_{\text{bol}, X}-L_{\text{bol}}$ relation for broad-line AGNs from \cite{dur20}.  }
 \label{fig:maiolino_plot}
\end{figure}

The upper limits of the X-ray luminosity provided in Table~\ref{tbl:chandra_upperlimit} suggest that the five {\it JWST} broad H$\alpha$ emitters do not exhibit the typical brightness levels expected for unobscured (type $1-1.9$) AGNs in the X-ray band.

According to the empirical relation between $L_{2-10~\text{keV}}$ and $L_{\text{H}\alpha, \text{broad}}$ for luminous Sloan Digital Sky Survey (SDSS) quasars, $\log L_{2-10~\text{keV}} = 0.83\log L_{\text{H}\alpha, \text{broad}} + 8.35$ \citep[in units of $\text{erg}~\text{s}^{-1}$;][]{jin12,yue24}, the expected X-ray luminosities of GLASS~160133, GLASS~150029, MSAID2008, MSAID4286, and MSAID38108 would be 
$L_{2-10~\text{keV}} =  (11.6, 8.67, 34.7, 219, 242)\times 10^{42}~\text{erg}~\text{s}^{-1}$ if these objects were unobscured AGNs.
The upper limits of $L_{2-10~\text{keV}}$ provided in Table~\ref{tbl:chandra_upperlimit} are below the expected X-ray luminosities, indicating that the X-ray to optical luminosity ratios in these objects are much smaller than those in typical unobscured AGNs.

Similarly, the upper limits of $L_{2-10~\text{keV}}$ can also be compared with the AGN bolometric luminosity estimated from the broad H$\alpha$ emission line luminosity
\citep{rei13,mat24}: 
\begin{eqnarray}
L_{\text{bol}} &=& 10.3 \times L_{5100\text{\AA}} \nonumber \\
&=& 10.3 \times 10^{44}~\text{erg}~\text{s}^{-1} \left(\frac{L_{\text{H}\alpha, \text{broad}}}{5.25\times 10^{42}~\text{erg}~\text{s}^{-1}}\right)^{\frac{1}{1.157}},
\label{eqn:lbol}
\end{eqnarray}
where $L_{5100\text{\AA}}$ is the monochromatic AGN continuum luminosity at $\lambda_{\text{rest}} = 5100$\AA, the prefactor of 10.3 is the optical bolometric correction factor \citep{ric06}, and the conversion between $L_{\text{H}\alpha, \text{broad}}$ and $L_{5100~\text{\AA}}$ is from Equation~2 of \cite{rei13}.
The bolometric luminosities for GLASS~160133, GLASS~150029, MSAID2008, MSAID4286, and MSAID38108 estimated from Equation~\ref{eqn:lbol} are $L_{\text{bol}} = (1.74, 1.28, 5.42, 36.9, 41.1) \times 10^{44}~\text{erg}~\text{s}^{-1}$, respectively.

As shown by \cite{mai24}, the X-ray bolometric correction factor $k_{\text{bol,X}} = L_{\text{bol}}/L_{2-10~\text{keV}}$ as a function of $L_{\text{bol}}$ serves as a diagnostic to distinguish X-ray normal and X-ray faint AGNs. 
The lower limits of $k_{\text{bol,X}}$ for these broad H$\alpha$ emitters are obtained as 58.5, 29.5, 82.4, 627, and 268 for $\Gamma=1.7$, and 49.4, 24.3, 64.6, 498, and 222 for $\Gamma=1.9$, respectively.
The lower limits of $k_{\text{bol,X}}$ for $\Gamma=1.7$ as a function of $L_{\text{bol}}$ are shown in Figure~\ref{fig:maiolino_plot}, along with the data points for {\it JWST} broad H$\alpha$ emitters taken from \cite{mai24} and \cite{ana24}\footnote{\cite{mai24} sample of 22 LRD and non-LRD broad H$\alpha$ emitters at $z>4$ is taken from their Table~1. \cite{ana24} sample is the following five LRD broad H$\alpha$ emitters from their Table 1, cross-matched with the spectroscopic sample with measured $L_{\text{H}\alpha, \text{broad}}$ provided in Table~3 of \cite{gre24}: ID = 1967 (MSAID4286), 8296 (MSAID13123), 8798 (MSAID13821), 28343 (MSAID38108), and 30782 (MSAID41225), removing secondary and tertiary images of the triply imaged lensed galaxy ID=8296. $L_{\text{bol}}$ is calculated from $L_{\text{H}\alpha, \text{broad}}$ using Equation~\ref{eqn:lbol}. Note that the lower limits of $k_{\text{bol,X}}$ for the same objects (MSAID4286 and MSAID38108) differ between this work and \cite{ana24}, primarily due to the different derivation methods used for the upper limits of the X-ray luminosity.}.
From the comparison with the empirical $k_{\text{bol,X}} - L_{\text{bol}}$ relation for unobscured AGNs \citep{dur20} as shown in Figure~\ref{fig:maiolino_plot}, these lower limits of $k_{\text{bol,X}}$ indicate that the broad H$\alpha$ emitters are X-ray faint compared to the known X-ray normal AGNs.

The estimated Eddington ratios of these broad H$\alpha$ emitters,  $L_{\text{bol}}/L_{\text{Edd}} = 0.81, 0.38, 0.90, 0.29, 0.15$, indicate that they are sub-Eddington accretors; thus, the empirically-known Eddington ratio-dependence of $k_{\text{bol,X}}$ \citep[e.g.,][]{dur20} is unlikely to be the primary reason for the X-ray weakness \citep[e.g.,][]{mai24}.
The X-ray weakness in the five LRD and non-LRD {\it JWST} broad H$\alpha$ emitters studied in this work is in line with what has been suggested in the literature for other non-LRD broad H$\alpha$ emitters \citep[][Figure~\ref{fig:maiolino_plot}]{mai24} and photometrically-selected LRDs and LRD broad H$\alpha$ emitters \citep[e.g.,][]{ana24,yue24,mai24,aki24} found in the various survey fields.

\subsection{Comparisons with the variability amplitudes of known AGNs}
\label{sec:comparison_with_know_AGNs}

\begin{figure}[tbp]
\center{
\includegraphics[clip, trim=0.0cm 0.0cm 0.0cm 0.0cm, width=3.4in]{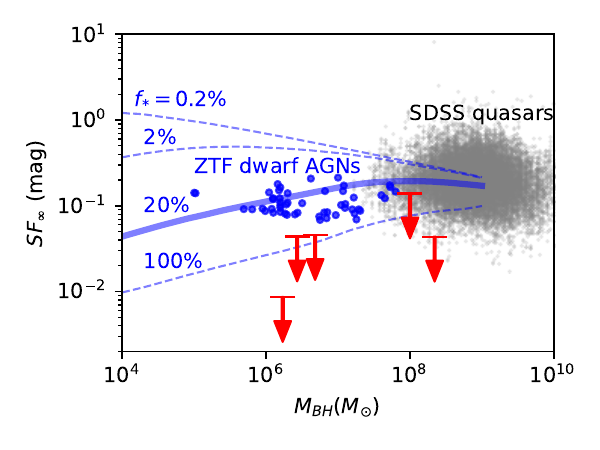}
}
 \caption{The asymptotic variability amplitude $\text{SF}_{\infty} = \sqrt{2}\sigma_{\text{d}}$ as a function of $M_{\text{BH}}$. The gray dots denote the SDSS Stripe82 five-band data ($u$, $g$, $r$, $i$, and $z$-band) of SDSS quasars \citep{mac10}, and blue points denote the ZTF $g$-band data of low-$z$ dwarf AGNs \citep{bur23}. The downward arrows indicate the conservative upper limits on $\text{SF}_{\infty}(4000\text{\AA})$ at the 90\% confidence level obtained for GLASS~160133, GLASS~150029, MSAID2008, MSAID4286, and MSAID38108 (from left to right; see Section~\ref{sec:likelihood}). The thick solid and thin dashed lines indicate the \cite{bur23}'s empirical $M_{\text{BH}}$-dependent model of $\text{SF}_{\infty}$, with $L_{\text{bol}}/L_{\text{Edd}} = 0.1$ and varying host dilution covering factors $f_{*} = 0.2\% - 100\%$.}
 \label{fig:burke_plot}
\end{figure}

The optical time-domain studies of the AGN continuum variability have revealed that the variability amplitude (characterized by the SF evaluated at infinity or the asymptotic variability amplitude in units of magnitude, $\text{SF}_{\infty}$) depends on the rest-frame wavelength, AGN luminosity, and $M_{\text{BH}}$ \citep[][and references therein]{bur23}:
\begin{eqnarray}
\log \text{SF}_{\infty} &=& A + B\log\left(\frac{\lambda}{4000\text{\AA}}\right) \nonumber\\
&+& C\left( M_{i} + 23 \right) + D\log \left(\frac{M_{\text{BH}}}{10^9~M_{\odot}}\right)
\label{eqn:variability_amplitude}
\end{eqnarray}
where $A = -0.51\pm0.02$, $B=-0.479\pm0.005$, $C=0.131\pm0.008$, and $D=0.18\pm0.03$.
$M_{i}$ is the absolute $i$-band magnitude of the AGN continuum \citep[approximately $M_{i} = 90 -2.5\log L_{\text{bol}}/\text{erg~s}^{-1}$;][]{bur23}.
By using $M_{i} \propto -2.5\log ( L_{\text{bol}}/L_{\text{Edd}} ) -2.5\log M_{\text{BH}}$, we can see $\log \text{SF}_{\infty} \propto B\log\left(\frac{\lambda}{4000\text{\AA}}\right) - 2.5C\log ( L_{\text{bol}}/L_{\text{Edd}} ) + (D-2.5C)\log M_{\text{BH}}$; thus, $\text{SF}_{\infty}$ is anti-correlated with $\log M_{\text{BH}}$ for a fixed Eddington ratio.

The host galaxy flux contamination dilutes the observed variability amplitude and reduces the value of $\text{SF}_{\infty}$ for low-luminosity AGNs \citep[e.g.,][]{kim20,bur23}.
By assuming an Eddington ratio of $L_{\text{bol}}/L_{\text{Edd}}=0.1$, $M_{\text{BH}} - M_{*}$ relation \citep[Equation~4 of][]{rei15}, and stellar mass-to-light ratio \citep[assuming a host galaxy color index of $g-r = 0.5$~mag;][]{zib09}, \cite{bur23} present empirical models of the host-diluted variability amplitude $\text{SF}_{\infty}$ based on Equation~\ref{eqn:variability_amplitude}, and compare the models with the observed values of $\text{SF}_{\infty}$ of the Sloan Digital Sky Survey Stripe82 quasars \citep{mac10} and the Zwicky Transient Facility (ZTF) dwarf AGNs at $z \approx 0.03$, as reproduced in Figure~\ref{fig:burke_plot}.
The \cite{bur23}'s models of the host-diluted variability amplitude $\text{SF}_{\infty}'$ are calculated as $\text{SF}_{\infty}' = \frac{L_{\text{AGN}}}{L_{\text{AGN}}+f_{*}L_{*}}\text{SF}_{\infty}$, where $L_{\text{AGN}}$ and $L_{*}$ are the AGN and host luminosities, and $f_{*}$ is the host dilution covering factor $f_{*}$ accounting for the fraction of the host galaxy luminosity enclosed in an aperture \citep[e.g., $f_{*} \approx 20\%$ for a $3''$-diameter aperture on the ZTF dwarf AGNs; Equation~11 of][]{bur23}.
Figure~\ref{fig:burke_plot} indicates that the unobscured AGNs from the SDSS and ZTF are adequately explained by the empirical model of $\text{SF}_{\infty}$, where the ZTF dwarf AGNs are intrinsically more variable but are more host-diluted thus exhibit the observed variability amplitudes comparable to or smaller than the luminous SDSS quasars.

There are several reasons to believe that the value of $f_{*}$ in the {\it JWST} broad H$\alpha$ emitters is not much greater than $20$\%.
The {\it JWST} broad H$\alpha$ emitters are believed to be hosting overmassive SMBHs compared to the local $M_{\text{BH}} - M_{*}$ relation \citep[e.g.,][]{har23,mai23,koc23,juo24,dur24}; thus, the $f_{*}$ factor should be intrinsically smaller than that assumed in the local dwarf AGNs.
Also, their high Eddington ratios imply that the host dilution covering factor $f_{*}$ in the {\it JWST} broad H$\alpha$ emitters would be effectively several times smaller than in the local dwarf AGNs accreting at sub-Eddington rates.
As mentioned in Section~\ref{sec:sample_selection}, the large EW of the broad H$\alpha$ emission line also suggests that the observed continuum is not likely to be dominated by the host galaxy's stellar light at least at the wavelengths around the $H\alpha$ line.
Especially for the LRDs, the SED model proposed by \cite{lab23} and \cite{gre24} assumes that the SED at $\lambda_{\text{rest}} \gtrsim 4000$\AA\ is dominated by the direct AGN emission (but see the caveats discussed in Section~\ref{sec:sample_lrd}), which suggests that the at least the F356W and F444W band photometry of MSAID2008, MSAID4286, and MSAID38108 can be assumed to be unaffected by the host galaxy dilution.
The UV-optical SED model for unobscured non-LRD broad H$\alpha$ emitters (like the objects in \citealt{har23}) adopted by \cite{mad24} in their Figure~2 assumes $\alpha_{\nu} \simeq -0.6$ and $+0.7$ ($f_{\nu} \propto \nu^{\alpha_{\nu}}$) for the AGN and host galaxy component, respectively, and the two components equally contribute to the emission at $\lambda_{\text{rest}}=1450$\AA.
In this model, the fraction of the host galaxy flux at $\lambda_{\text{rest}}=4000$\AA\ is only about 20\%.
Based on these considerations, the host galaxy flux contamination is expected to be not so large as to hinder the variability signals, and a variability amplitude of $\gtrsim 0.1$~mag should be detectable if the {\it JWST} broad H$\alpha$ emitters are AGNs.

Nevertheless, as shown in Figure~\ref{fig:burke_plot}, the upper limits on the variability amplitudes of the {\it JWST} broad H$\alpha$ emitters at $\lambda_{\text{rest}} = 4000$\text{\AA} indicate that their variability amplitude is much less than $0.1$~mag.
Remarkably, comparisons with the \cite{bur23}'s models of the host-diluted variability amplitude in Figure~\ref{fig:burke_plot} suggest that the host galaxy dilution effect alone cannot explain the small variability amplitudes of {\it JWST} broad H$\alpha$ emitters (except for MSAID4286) even when the maximal host dilution ($f_{*} \simeq 100\%$) is assumed.
We note that the non-detection of the photometric variability in the long wavelength bands ($\lambda_{\text{rest}} > \lambda_{\text{H}\alpha}$) implies that the dust reddening cannot be the reason for the small variability amplitudes.

From these comparisons, we conclude that the {\it JWST} broad H$\alpha$ emitters do not exhibit the expected flux variations as deduced from the known AGNs.
This suggests that the {\it JWST} broad H$\alpha$ emitters cannot be explained by the standard AGN scenario as proposed in the literature.
The term `standard' is conservatively used here as assuming certain 'non-standard' AGN structures/geometries might still allow the AGN model to remain consistent with the non-detections of the photometric variability and broad-band SEDs of the {\it JWST} broad H$\alpha$ emitters.
The possible AGN and non-AGN models for the {\it JWST} broad H$\alpha$ emitters are discussed below.

\subsection{Possible models for the {\it JWST} broad H$\alpha$ emitters}
\label{sec:possible_alternative}

As concluded in the previous section, the non-detection of the photometric variability disfavors the standard type 1-1.9 AGN model for both of the LRD and non-LRD broad H$\alpha$ emitters.
Since the five objects analyzed in this study are typical examples of the LRD and non-LRD broad H$\alpha$ emitters, the non-variability observed in these objects suggests that similar behavior may be present in the general population of the {\it JWST} broad H$\alpha$ emitters. 
However, we note that the small sample size and possible sample selection biases should be kept in mind when extrapolating to the broader population.

As mentioned earlier (Section~\ref{sec:xray_weakness}), the observational fact that X-ray emission has not been detected in the {\it JWST} broad H$\alpha$ emitters also suggests that they are not standard AGNs \citep[see also][]{ana24,koc24,mai24,aki24}.
Based on the AGN scenario, in Compton-thick AGNs where rest-frame hard X-rays are completely obscured, the BLR emission is generally also obscured; thus, the visibility of the broad H$\alpha$ emission line is puzzling.

\begin{figure*}[tbp]
\center{
\includegraphics[clip, trim=0.0cm 0.0cm 0.0cm 0.0cm, width=7.0in]{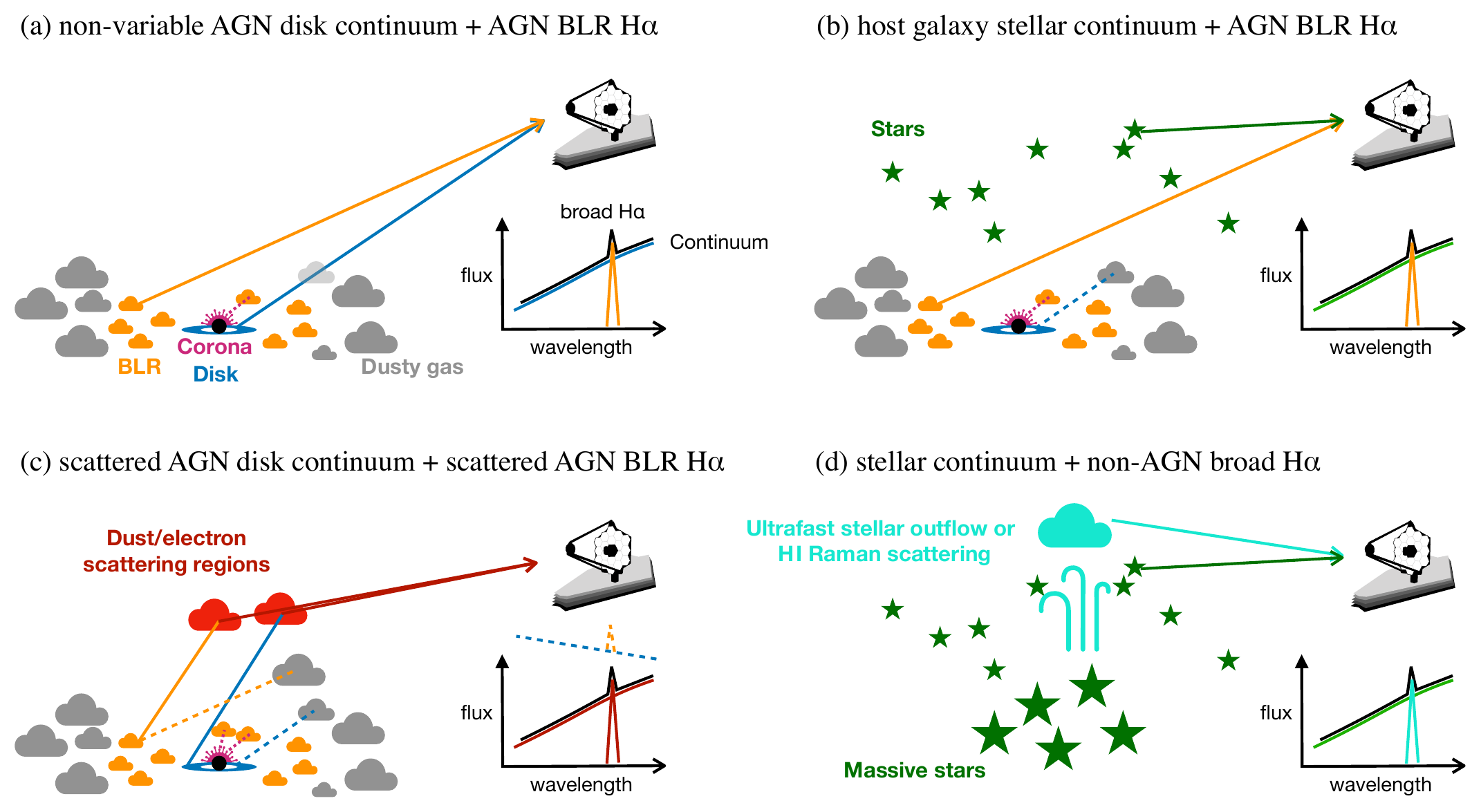}
}
 \caption{Schematic pictures of possible scenarios for the non-variable rest-frame optical continuum and broad H$\alpha$ emission line in the {\it JWST} broad H$\alpha$ emitters.
 The insets within the figure show an outline of the optical continuum + broad H$\alpha$ spectral components (color-coded). 
 Scenarios (a), (b), and (c) require an AGN, while scenario (d) is a non-AGN model (Section~\ref{sec:possible_alternative}). 
 We require the {\it JWST} broad H$\alpha$ emitters to be X-ray faint (Section~\ref{sec:xray_weakness}).
 (a): the continuum and broad H$\alpha$ line are (moderately-reddened) direct AGN emissions, but unlike typical AGNs, they do not exhibit flux variability for some reason. Compton-thick materials (e.g., BLR gas clouds) hide the X-ray corona, but do not obscure the accretion disk continuum.
 (b): the AGN disk continuum is dust-obscured but the AGN BLR emission is unobscured due to some specific geometry of the dusty materials, and the observed continuum is host stellar light-dominated. Compton-thick materials (e.g., BLR gas and dusty clouds) hide the X-ray corona.
 (c): The observed spectrum is dominated by the scattered AGN disk continuum and scattered BLR H$\alpha$ emission line, and the direct AGN emission is obscured. The scattered emission can be moderately reddened due to dust extinction inside the scattering regions. Compton-thick materials with a high covering factor hide the X-ray corona from both the observer and the scattering regions. This model requires an intrinsically super-luminous AGN embedded inside the dusty materials (as indicated by dashed lines in the inset). 
 (d): the continuum is the stellar light and the broad H$\alpha$ emission line is either from ultrafast and dense/low-metallicity stellar outflows or inelastic Raman scattering of stellar UV continua by neutral hydrogen atoms. This model is intrinsically X-ray faint unless there are ultraluminous stellar X-ray sources.}
 \label{fig:schematic}
\end{figure*}

These observational facts might lead one to conclude that the {\it JWST} broad H$\alpha$ emitters are non-AGNs.
However, if we consider physical conditions and complex geometries beyond the standard AGN model, there could be possibilities that the broad H$\alpha$ emission line and UV-optical continuum of the broad H$\alpha$ emitters are contributed by some form of AGN emission, while simultaneously obscuring the AGN X-ray emission.
Several theoretically possible non-standard AGN models for the {\it JWST} broad H$\alpha$ emitters and their caveats are discussed below (scenarios a, b, and c in Figure~\ref{fig:schematic}).
Then, we discuss the alternative non-AGN scenario that might explain the nature of the broad H$\alpha$ emitters (scenario d in Figure~\ref{fig:schematic}).

\subsubsection*{\MakeLowercase{(a)} Intrinsically non-variable AGN accretion disk continuum}
\label{sec:nonvariable_AGN_disk}

Since the AGN variability is stochastic, it is conceivable that the AGN variability could not occur by chance. 
However, the fact that the flux variations were not detected in all of the LRD and non-LRD broad H$\alpha$ emitters studied here leads us to infer that a significant proportion of the {\it JWST} broad H$\alpha$ emitters identified in the literature does not exhibit the variability typically seen in normal AGNs.

It may be possible to consider that AGNs in the {\it JWST} broad H$\alpha$ emitters are in a different state from the known AGNs and intrinsically non-variable ones.
As previous works have shown that the AGN continuum variability at a fixed rest-frame UV-optical wavelength is independent of the redshift, i.e., there is no cosmological evolution in the AGN accretion disk physics \citep[e.g.,][]{sub21}, the high redshift of the {\it JWST} broad H$\alpha$ emitters alone is unlikely to be the reason for the non-detection of the photometric variability.
The mass accretion rates of the {\it JWST} broad H$\alpha$ emitters are typically higher than those of known AGNs, which may lead to different variability properties\footnote{After the submission of this manuscript to arXiv and ApJ, a paper by \cite{ina24} appeared on arXiv, proposing that photon trapping in super-Eddington accretion disks could suppress variability signals because photon trapping weakens the luminosity response to fluctuations in the mass accretion rate \citep[e.g.,][]{wat00}.}.
Although the direct view of the accretion disk continuum seems to suggest that the AGN coronal X-ray emission should also be directly observable (contrary to the observational facts), it is possible to suppress the X-ray flux by supposing that compact dust-free Compton-thick gas clouds, such as disk outflows or BLR clouds, are present along the line of sight, by which the coronal X-ray emission is obscured whereas the UV-optical accretion disk emission is directly observable (see Figure~\ref{fig:schematic}).
It should also be noted that local Narrow-line Seyfert 1 (NLSy1) Galaxies \citep[believed to have high Eddington ratios and thought to be local analogs of the {\it JWST} broad H$\alpha$ emitters;][]{mai24} are known to have steeper X-ray spectra than normal AGNs, which may result in an X-ray weakness in the hard X-ray bands.

However, the mass accretion rates of the {\it JWST} broad H$\alpha$ emitters are not expected to be as high as super-Eddington rates \citep[][see Section~\ref{sec:xray_weakness}]{koc23,har23,mai24}, and the disk state transition from the standard accretion disk \citep{sha73} to slim-disk \citep{abr88} is not anticipated.
Given that the broad H$\alpha$ emitters are in the same standard accretion disk state as known AGNs, there is no reason to expect completely different variability characteristics between the two populations.
Even if the accretion disks in the broad H$\alpha$ emitters are partially in the slim-disk state at their inner disk regions, it is unclear if the slim disks are more stable than standard accretion disks and exhibit much less variability, as observed \citep[from the perspective of the local/global stability of the accretion disk; e.g.,][]{szu96,jiang19}.
Observationally, local NLSy1 galaxies have been observed to exhibit UV-optical variability comparable to normal broad-line AGNs \citep{rak17}.
The scenario attributing the variability non-detection to the non-standard disk state is not theoretically/observationally motivated, but it should be carefully tested through future observations of broad-band SEDs and various emission lines.

\subsubsection*{\MakeLowercase{(b)} Disk continuum-obscured, BLR-unobscured AGNs}
\label{sec:continuum_obscured_AGN}

Considering a specific geometry in which the AGN accretion disk continuum is heavily obscured while the AGN BLR emission lines remain unobscured, the presence of the AGN can be reconciled with the non-detection of photometric variability.
In this scenario, the broad H$\alpha$ emission line is the direct AGN BLR emission, and the observed continuum is the non-AGN (stellar) origin.
Since the solid angle subtended by the AGN accretion disk is much smaller than that of the AGN BLR, such obscuring geometry is theoretically possible if compact dusty (optically thick) clouds are present along our line of sight to the AGN.
Indeed, such a geometry, often referred to as the clumpy dusty torus, has been widely recognized as a viable model to explain the observed properties of nearby AGNs \citep[e.g.,][]{hon06,nen08b,hon10}.
The same obscuring materials would also cause the extinction of the X-ray emission from the AGN corona.
This scenario may explain the high EWs of the broad H$\alpha$ emission line observed in a fraction of the {\it JWST} broad H$\alpha$ emitters \citep{mai24}.
Moreover, the stellar origin of the rest-frame optical continuum is favored from the rest-frame NIR observations of LRDs ---mostly photometrically-selected LRDs, with several LRD broad H$\alpha$ emitters--- which reveal that the observed NIR color with a bump at $\lambda_{\text{rest}}\sim 1.6\mu\text{m}$ is more consistent with a moderately obscured stellar continuum than the AGN hot dust continuum \citep{wil24,per24,aki24}.

The problem with this scenario is the lack of known AGNs requiring such a specific obscuring geometry.
The optical variability studies of the SDSS quasars indicate that essentially all the quasars show some level of flux variability on timescales of several months-to-years \citep[e.g.,][]{ses07}, suggesting that the continuum-obscured BLR-unobscured AGNs, if they exist, would be extremely rare among the known broad-line AGNs.
Based on insights from previous AGN spectroscopic/SED studies, we can at least say that it seems unlikely that the continuum-obscured BLR-unobscured AGNs constitute a majority among the entire AGN population.
Moreover, if this scenario were correct, there should also be a similar number of type 1 (continuum-unobscured/BLR-unobscured) AGNs with the {\it JWST} broad H$\alpha$ emitters (continuum-obscured/BLR-unobscured) when considering that viewing angles are random. 
However, such numerous type 1 AGNs have not been found in any survey, including those conducted with {\it JWST}.

To directly verify/disprove this scenario observationally, multi-epoch spectroscopic data are needed to examine the flux variability in the broad H$\alpha$ emission lines, which could be feasible with future {\it JWST}/NIRSpec follow-up spectroscopy.
Also, detecting the Balmer break and/or stellar absorption features in the rest-frame optical spectrum through deep spectroscopy will lead to the direct confirmation of the scenario in which the continuum emission is of stellar origin \citep[e.g.,][]{wan24}.

\subsubsection*{\MakeLowercase{(c)} Scattering-dominated AGN disk and BLR emission}
\label{sec:scattered_AGN_BLR}

One scenario proposed to explain the blue excess at $\lambda_{\text{rest}} < 3000\text{\AA}$ in LRDs is the unobscured scattered AGN accretion disk emission \citep{lab23,gre24}.
For MSAID4286, \cite{gre24} present an SED model composed of direct and scattered AGN components in their Fig.~6, in which a reddened ($A_{V}=2.7$~mag) direct AGN emission is fitted to the rest-frame optical emission at $\lambda_{\text{rest}} > 4000\text{\AA}$ and an unobscured scattered AGN component with a scattered light fraction of $f_{\text{scat}}=2.5$\% is fitted to the rest-frame UV continuum at $\lambda_{\text{rest}} < 3000\text{\AA}$.
This model for the LRDs itself is disfavored by the non-detection of the photometric variability because, if the model were correct, the direct AGN emission at $\lambda_{\text{rest}} > 4000\text{\AA}$ should be variable.

However, if the rest-frame optical continuum at $\lambda_{\text{rest}} > 4000\text{\AA}$ and broad H$\alpha$ emission line are also assumed to be scattered components of the unseen heavily obscured AGN, and if the scatterers are extended and located far from the AGN region, this additional scattering would produce a non-variable optical emission since the superposition of scattered light from multiple regions dilutes the AGN luminosity variations.
For LRDs, the two scattering regions required to produce the blue UV excess and red optical continuum need to be distinct; they have different extinction ($A_{V} \sim 0$~mag and $\sim 2.7$~mag, respectively) and scattering efficiencies (the latter should have a much higher scattered light fraction than the former).
For non-LRD broad H$\alpha$ emitters (e.g., GLASS~160133 and GLASS~150029), a single scattering region of a high scattering efficiency may be enough to produce the observed continuum and broad H$\alpha$ emission line.
The observational fact of higher electron density in the ISM of higher-$z$ galaxies \citep[e.g.,][]{iso23} may support the assumption of the regions with highly efficient electron scattering in the {\it JWST} broad H$\alpha$ emitters.

The main issue with adopting this model for the {\it JWST} broad H$\alpha$ emitters is that it requires intrinsically very luminous heavily-obscured AGNs and/or unreasonable high scattering efficiencies.
If we adopt the scattered broad H$\alpha$ emission line model, since the Eddington accretion rates inferred from the observed H$\alpha$ luminosities are already close to the Eddington limit \citep[][see Section~\ref{sec:xray_weakness}]{koc23,har23,mai24}, the intrinsic broad H$\alpha$ luminosities would translate to super-Eddington luminosities of $> 10~L_{\text{Edd}}$, given that the scattered light fraction can only reach at most a few percent \citep[e.g.,][]{gre24}.
In practice, due to photon trapping effects, the AGN accretion disk luminosity is not expected to exceed a few times $L_{\text{Edd}}$ even if a super-Eddington mass accretion rate is achieved \citep[e.g.,][]{ohs02}.
In addition, to date, no known type $1-1.9$ AGN (other than {\it JWST} broad H$\alpha$ emitters) requires an AGN scattering-dominated UV-optical SED model (the UV-optical emission of essentially all of the known broad-line AGNs is consistent with the direct AGN emission; see Section~\ref{sec:continuum_obscured_AGN}).

Moreover, such scattering regions with high scattering efficiencies would also produce a measurable amount of scattered X-ray emission in the X-ray bands, which is likely to contradict the X-ray weakness of the {\it JWST} broad H$\alpha$ emitters.
To avoid this, Compton-thick materials with a high covering factor need to be introduced ad hoc, enveloping not only the line-of-sight toward the observer but also the entire X-ray corona \citep[see][and Figure~\ref{fig:schematic}]{mai24}.

Although this scenario invoking the scattered AGN BLR broad H$\alpha$ emission is probably inappropriate as an explanation for the majority of the {\it JWST} broad H$\alpha$ emitters, it may account for a minority of them.
Through deep mid-IR imaging by {\it JWST}/MIRI, it will be possible to confirm the presence or absence of the direct dust torus emission from an intrinsically bright obscured AGN assumed in this scenario \citep[see, e.g.,][]{wil24,per24,wan24,aki24,ian24}.

\subsubsection*{\MakeLowercase{(d)} Non-AGN broad H$\alpha$ emission mechanisms: fast/dense galactic outflows or hydrogen Raman scattering}
\label{sec:nonAGN_scenario}

Finally, we consider the possibility that the {\it JWST} broad H$\alpha$ emitters are galaxies that do not contain AGN.
As mentioned above (Section~\ref{sec:possible_alternative}(b)), rest-frame NIR observations of LRDs indicate that the NIR continuum is not as red as the dust emission from AGN hot dust, suggesting a stellar continuum origin for the observed optical-NIR continuum in LRDs and broad H$\alpha$ emitters in general \citep{wil24,per24}.
Below, we discuss several possible non-AGN mechanisms for the broad H$\alpha$ emission line.

In the {\it JWST} the broad H$\alpha$ emitters, the broad lines with a velocity width of $\gtrsim 1,000$~km~s$^{-1}$ are exclusively observed in the permitted H$\alpha$ (and sometimes H$\beta$), with no corresponding broad component observed in forbidden [\ion{O}{3}] lines.
Suppose the broad H$\alpha$ emission line is interpreted as the emission from fast-moving gas clouds.\footnote{After the submission of this manuscript to arXiv and ApJ, the paper by \cite{bag24} appeared on arXiv, proposing a scenario in which the extremely high stellar densities as observed in LRDs influence gas dynamics and may contribute to the generation of high-velocity-dispersion gas clouds and, consequently, broad emission lines. This represents an intriguing alternative to fast, dense galactic outflows; however, explaining broad lines with FWHM $\gtrsim$ 3500 km s$^{-1}$, as observed in {\it JWST} broad H$\alpha$ emitters \citep[e.g.,][]{har23,gre24,mai24}, solely through this mechanism would likely be challenging.}
In that case, the absence of the forbidden lines indicates that the gas is very dense with an electron number density of $n_{e} \geq 10^{5-6}~\text{cm}^{-3}$, or has a very low-metallicity of $Z/Z_{\odot} < 0.01$.
The detection of the strong narrow [\ion{O}{3}] lines indicates that the galaxies' ISM is metal-enriched, and it is unreasonable to consider the presence of very low-metallicity galactic outflows in these systems.
Also, such high particle densities are quite different from normal stellar winds or galactic outflows \citep{gre24,mai24}.
Besides the AGN BLR, no persistent astrophysical line-emitting gaseous environments of such high velocity and high density are currently known.

The {\it JWST} broad H$\alpha$ emission lines may originate from individual luminous transients (e.g., SNe, LBVs, and TDEs) in the star-forming galaxies \citep[e.g.,][]{izo07,gus24,mai24}, but the non-detection of the NIRCam photometric variability implies that this possibility is unlikely at least for the objects studied here.
Furthermore, the event rate of such luminous transients necessary to account for the observed high luminosity broad H$\alpha$ line would be too low to be observed in a significant number of the high-$z$ low-mass galaxies of interest \citep[$M_{*}\sim 10^{9}~M_{\odot}$;][]{har23}.
An ensemble of multiple SNe might give rise to a measurable amount of persistent broad H$\alpha$ emission line \citep[e.g.,][]{izo07}, but it would require an unreasonably high star-formation rate for the high-$z$ low-mass galaxies \citep[e.g.,][]{juo24}.
Moreover, the multiple SN remnants/superbubbles would produce broad components not only in the hydrogen Balmer lines but also in the [\ion{O}{3}] forbidden lines \citep[e.g.,][]{roy92,izo07}, which are inconsistent with the spectroscopic properties of the {\it JWST} broad H$\alpha$ emitters.
The lack of velocity offset between the narrow and broad emission line components, and the symmetry of the line profiles also suggest that they are not due to usual stellar-driven outflows \citep{wan24}.

Rather than interpreting the broad H$\alpha$ line as Doppler-broadened emission, we propose an alternative possible explanation: that it could originate from optical photons produced through the Raman scattering of UV continuum photons by neutral hydrogen atoms.
When UV continuum photons (of stellar origin) around the Ly$\beta$ resonance wavelength are scattered by a hydrogen atom in the ground $1s$ state, a certain fraction of the outgoing photon is inelastically down-scattered into optical wavelengths around the H$\alpha$ wavelength by leaving the hydrogen atom in the $2s$ excited state, forming a broad H$\alpha$ emission feature \citep[e.g.,][and references therein]{lee00,kok24}.
The same Raman conversion occurs between the Ly$\gamma$ and H$\beta$ resonances, and henceforth.
The line profile of the Raman-scattered broad emission feature is not relevant to the kinematics of the gas but is determined by the scattering opacity (= scattering cross-section of the hydrogen atom multiplied by the hydrogen column density $N_{\text{H}}$).
This Raman scattering feature has been observed in both Galactic and extragalactic \ion{H}{2} regions associated with very young O-type stars \citep{dop16,hen21}, and possibly in a local blue compact dwarf galaxy SBS~0335-052E \citep{hat23}.

The Raman-scattered H$\alpha$ emission feature would be as broad as $>$ 1000~km~s$^{-1}$ if there is a hydrogen gas of $N_{\text{H}} > 10^{20}~\text{cm}^{-2}$ surrounding strong UV sources \citep[Equation~32 of][]{kok24}.
Although the Raman-scattered H$\alpha$ emission line profile from a uniform-density gas is flat-top with a Lorenzian wing that does not resemble the observed line profile, the gross H$\alpha$ emission line profile from multiple emission regions and scattering regions with different $N_{\text{H}}$ can have an arbitrary shape.

The explanation of the broad H$\alpha$ emitters through the hydrogen Raman scattering may face a limitation due to the absence of the expected intense narrow H$\alpha$ emission line from the ionized gas where the Raman scattering occurs. 
In other words, the observed flux ratio between narrow and broad emission lines might be smaller than anticipated \citep[see, e.g.,][]{dop16}.
The rich hydrogen gas in these {\it JWST}-detected high-$z$ star-forming galaxies \citep[evident as Ly$\alpha$ absorbers, and occasionally observed as the H$\alpha$ absorbers; e.g.,][]{mat24,mai24} may be responsible for the production of Raman-scattered broad Balmer features \citep{dop16,kil23,kok24}, while simultaneously absorbing the narrow core emission of the H$\alpha$ line and reducing the contrast between the narrow and broad line fluxes.
Further observations, such as deep spectroscopic searches for broad emission line components of hydrogen, helium, and other elements, are needed to verify/refute this Raman-scattering scenario.
The relationship between the emergence of broad H$\alpha$ emission and peculiar continuum SED shape seen in the {\it JWST}-detected high-$z$ galaxy sample is also an intriguing point for which further investigation is needed\footnote{After the submission of this manuscript to arXiv and ApJ, \cite{juo24b} was posted on arXiv. Their argument is based on the observation that the multiple hydrogen broad emission lines (H$\alpha$, H$\beta$, and Pa$\beta$) in the bright low-$z$ LRD GN-28074 ($z = 2.26$) exhibit nearly identical velocity widths, which they claim conflicts with \cite{kok24}, where broad lines produced by hydrogen Raman scattering are predicted to have significantly different widths for different hydrogen lines. Investigating whether the same behavior (i.e., multiple hydrogen broad emission lines exhibiting similar velocity widths) occurs in other, more typical LRDs and non-LRDs broad H$\alpha$ emitters remains an urgent observational task.}.
Once confirmed, the broad Raman-scattering feature may provide us with a unique tool to investigate the otherwise-unseen stellar UV radiation field and spatial extent of the atomic hydrogen in these high-$z$ star-forming galaxies \citep{hen21}.

As mentioned in Section~\ref{sec:intro}, confirming the non-AGN scenario for the broad H$\alpha$ emitters entails several implications.
The non-AGN scenario implies that the spatial density of the broad-line AGNs at $z \gtrsim 4$ is not necessarily as high as suggested in the literature, and it alleviates the need for an unexpectedly abundant population of highly-accreting low-mass AGNs in the high-$z$ Universe.
Both of the LRD and non-LRD broad H$\alpha$ emitters are intrinsically X-ray weak, which suggests, under the AGN scenario, that they are Compton-thick AGNs \citep{yue24,ana24,mai24}.
The X-ray non-detection is not surprising if they are not AGNs but star-forming galaxies; the current upper limits on the rest-frame X-ray luminosity still allow the presence of X-ray emission from the host stellar populations.
We point out that the non-AGN scenario for the LRD and non-LRD broad H$\alpha$ emitters alleviates the potential tension between the expected cumulative AGN X-ray radiation field and observed unresolved X-ray background radiation \citep{pad23,mai24}.

\section{Summary and conclusions}
\label{sec:conclusions}

We examined the rest-frame UV-optical variability of five {\it JWST}/NIRSpec broad H$\alpha$ emitters in the Abell 2744~field, with the multi-epoch {\it JWST}/NIRCam imaging data taken with the six wide photometric bands (F115W, F150W, F200W, F277W, F356W, and F444W).
Three of the objects are classified as LRDs, whereas the remaining two are classified as non-LRDs \citep{har23,gre24}.
Under the assumption that the broad H$\alpha$ emission line originates from the AGN BLR, the BH masses of these objects are in a range of $M_{\text{BH}} = 10^{6-8}~M_{\odot}$.
The rest-frame temporal sampling interval of the NIRCam data ($\Delta t \sim 400-500$~days$/(1+z)$) is comparable to the decorrelation time scales of the putative AGNs ($\tau_{\text{d}} \sim 20-100$~days); thus, the flux variations should be detectable if the AGNs were present.

We did not detect any variability in any of the bands, and the obtained upper limits on the DRW asymptotic variability amplitude are inconsistent with that observed in known AGNs even if the significant host galaxy flux contamination is considered. 
This result suggests that a large fraction of the LRD and non-LRD broad H$\alpha$ emitters reported in the literature is not to be classified as standard AGNs.
The conclusion that most of the LRD and non-LRD broad H$\alpha$ emitters are not standard AGNs aligns with the observational fact that these populations are X-ray faint, confirmed by the deep {\it Chandra} observations.

The broad emission line component observed in the broad H$\alpha$ emitters is not seen in the forbidden [\ion{O}{3}] line, which requires a very high particle density of $n_{e} \geq 10^{5-6}~\text{cm}^{-3}$ and/or a very low-metallicity of $Z/Z_{\odot} < 0.01$ if the broad component is interpreted as a Doppler-broadened emission line from fast-moving gas clouds.
Such high density and low-metallicity are inconsistent with normal stellar winds or galactic outflows, but the possibility of the existence of the fast and dense/low-metallicity outflow phenomenon unique to high-$z$ low-metallicity galaxies cannot be ruled out.
The broad H$\alpha$ emission line might originate from individual transients (e.g., SNe and LBVs), but the non-detection of the NIRCam photometric variability implies that this possibility is unlikely at least for the the objects studied here.

We have considered several non-standard AGN structures that could explain the broad H$\alpha$ emitters: ($a$) an intrinsically non-variable AGN accretion disk continuum, ($b$) a host galaxy-dominated continuum, and ($c$) scattering-dominated AGN emission.
Considering the challenges in physically realizing these non-standard AGN structures, we have also proposed non-AGN models ($d$) in which the broad H$\alpha$ emission line could be a signature of unusually fast and dense/low-metallicity star-formation-driven outflows, or Raman scattering of the stellar UV continuum photons by hydrogen atoms in the galaxy (Figure~\ref{fig:schematic}).
We noted that such a signature of Raman scattering is observed in Galactic and extragalactic \ion{H}{2} regions of young O-type star associations.
The relationship between the emergence of the broad H$\alpha$ emission line (produced via Raman scattering or other mechanisms) and the peculiar $v$-shaped SED observed in LRDs is unclear and warrants further investigation.

It should be noted that the objects studied in this work constitute a subset of the known {\it JWST} broad H$\alpha$ emitters, and we do not claim that all the broad H$\alpha$ emitters identified in the literature share the same characteristics as the objects studied here.
A part of the known broad H$\alpha$ emitters exhibits unmistakably evident AGN signatures.
For example, MSAID45924 in the sample of \cite{gre24} exhibits not only the very broad H$\alpha$ line of $\text{FWHM} = 4500$~km~s${}^{-1}$ but also the high ionization [\ion{Ne}{5}]$\lambda 3426$ line (ionization potential of 95~eV); thus, it can be identified as an AGN with confidence \citep[see also GS~3073;][]{ubl23}.
Nevertheless, the non-detection of the photometric variability in all the {\it JWST} broad H$\alpha$ emitter samples studied here suggests that many known LRD and non-LRD broad H$\alpha$ emitters are similarly non-variable and, thus, unlikely to be standard type $1-1.9$ AGNs.
Further {\it JWST} observations, such as the spectroscopic search for the rest-frame optical stellar absorption features and MIR imaging search for the hot dust emission, will be able to test the various scenarios explaining the {\it JWST} broad H$\alpha$ emitters.

In this study, we focused on analyzing only the five objects around Abell~2744 for which multi-band multi-epoch NIRCam data are available in order to obtain the tight upper limits on the rest-frame optical variability amplitude.
However, there are many more objects in the Abell~2744 and other survey fields with single-band or two-band NIRCam and NIRISS multi-epoch data, and we will examine the variability of these other objects in the forthcoming paper.
Also, accumulating multi-epoch {\it JWST} NIR-MIR imaging/spectroscopic data in the {\it JWST} North Ecliptic Pole Time-Domain Field \citep[NEP-TDF; e.g.,][]{jansen18,obr24,jha24} and other deep survey fields (e.g., GOODS and COSMOS) will be beneficial to firmly identify unobscured/(mildly-)obscured AGN population in the high-$z$ Universe through the variability detection.
Combinations of multi-wavelength time-domain datasets produced from {\it JWST} and future surveys/missions (e.g., Vera C. Rubin Observatory, {\it Euclid}, {\it SPHEREx}, and {\it Roman}) will enable us to perform a variability-based selection for AGNs with various obscuration levels in a wide luminosity range.


\section*{Acknowledgments}

This work was supported by JSPS KAKENHI Grant Numbers 24K17097 and 24H00245.
The authors thank Kohei Ichikawa, Kohei Inayoshi, and Masami Ouchi for fruitful discussions.

This work is based on observations made with the NASA/ESA/CSA James Webb Space Telescope. 
The data were obtained from the Mikulski Archive for Space Telescopes at the Space Telescope Science Institute, which is operated by the Association of Universities for Research in Astronomy, Inc., under NASA contract NAS 5-03127 for JWST. 
These observations are associated with programs \#1324, \#2561, \#2756, \#3516, and \#3990.
All the {\it JWST} data used in this paper can be found in MAST: \dataset[10.17909/hj9a-cr40]{http://dx.doi.org/10.17909/hj9a-cr40}.
The authors acknowledge the UNCOVER and GLASS teams for developing their observing program with a zero-exclusive-access period.

This research has made use of data obtained from the {\it Chandra} Data Archive, and software provided by the {\it Chandra} X-ray Center (CXC) in the application packages CIAO and Sherpa.
This paper employs a list of {\it Chandra} datasets, obtained by the {\it Chandra} X-ray Observatory, contained in the {\it Chandra} Data Collection (CDC) 464~\dataset[doi:10.25574/cdc.464]{https://doi.org/10.25574/cdc.464},
including observations with principal investigators L.~David, G.~Garmire, J.~Kempner, and \'{A}.~Bogd\'{a}n.


%

\vspace{5mm}
\facility{
JWST (NIRCam, NIRSpec),
CXO (ACIS),
MAST
}


\software{
Astropy~v5.1 \citep{2013A&A...558A..33A,2018AJ....156..123A,2022ApJ...935..167A},
Matplotlib~v3.6.2 \citep{matplotlib07},
CIAO v4.15 \citep{fru06},
Sherpa~v4.15.0 \citep{sherpa22},
XSPEC~v12.13.1 \citep{arn96},
Photutils~v1.6.0 \citep{larry_bradley_2023_7946442},
image\_registration~v0.2.10 \citep{gin14}, 
image1overf.py~v2023-07-17, 
JWST Science Calibration Pipeline \citep{jwst_v1.14.0},
WebbPSF \citep{per14},
ChatGPT (OpenAI, 2024)
}



\appendix

\section{Log of the {\it JWST}/NIRCam wide-band observations}
\label{sec:log_of_observation}

\startlongtable
\begin{deluxetable*}{ccccccc}
\tablecolumns{7}
\tablecaption{Log of the {\it JWST}/NIRCam wide-band observations \label{tbl:observations}}
\tablehead{
  \colhead{Band} & 
  \colhead{Start Time} &
  \colhead{End Time} &
  \colhead{Exposure time} &
  \colhead{MJD-mid} &
  \colhead{$\Delta t_{\text{obs}}$}&
  \colhead{Program~ID}\\
  \colhead{} & 
  \colhead{(UTC)} &
  \colhead{(UTC)} &
  \colhead{(second)} &
  \colhead{(days)} &
  \colhead{(days)} &
  \colhead{}
  }
\startdata
  \multicolumn{7}{c}{ GLASS~160133, GLASS~150029 } \\
  \multicolumn{7}{c}{ epoch~1 } \\
  F115W  & 2022-11-02T08:27:04.089 & 2022-11-04T18:15:08.372 & 20807.9 & 59886.53459 & 0.16869 & 2561 \\
  F150W  & 2022-11-02T10:37:52.658 & 2022-11-04T19:39:36.207 & 20807.9 & 59886.60888 & 0.24298 & 2561 \\
  F200W  & 2022-11-02T12:48:19.786 & 2022-11-04T16:49:36.153 & 13399.5 & 59886.61761 & 0.25171 & 2561 \\
  F277W  & 2022-11-02T08:27:04.153 & 2022-11-04T12:17:25.481 & 12562.0 & 59886.36590 & 0.00000 & 2561 \\
  F356W  & 2022-11-02T10:37:52.722 & 2022-11-04T14:26:05.281 & 12562.0 & 59886.45595 & 0.09005 & 2561 \\
  F444W  & 2022-11-02T14:59:08.410 & 2022-11-04T19:39:36.207 & 16491.7 & 59886.72250 & 0.35660 & 2561 \\
  \multicolumn{7}{c}{ epoch~2 } \\
  F115W  & 2023-08-01T09:51:56.306 & 2023-08-01T11:26:14.683 &  5024.8 & 60157.44377 & 271.07787 & 2561 \\
  F150W  & 2023-08-01T11:32:08.909 & 2023-08-01T13:06:38.037 &  5024.8 & 60157.51351 & 271.14761 & 2561 \\
  F200W  & 2023-08-01T13:12:43.015 & 2023-08-01T14:47:33.583 &  5024.8 & 60157.58337 & 271.21747 & 2561 \\
  F277W  & 2023-08-01T09:51:56.370 & 2023-08-01T11:26:14.619 &  5024.8 & 60157.44377 & 271.07787 & 2561 \\
  F356W  & 2023-08-01T11:32:08.972 & 2023-08-01T13:06:37.973 &  5024.8 & 60157.51351 & 271.14761 & 2561 \\
  F444W  & 2023-08-01T13:12:43.015 & 2023-08-01T14:47:33.583 &  5024.8 & 60157.58337 & 271.21747 & 2561 \\ 
  \multicolumn{7}{c}{ epoch~3 } \\
  F356W  & 2023-12-01T18:00:32.955 & 2023-12-10T11:17:31.518 &  3156.6 & 60284.11092 & 397.74502 & 3516 \\ \hline
  \multicolumn{7}{c}{ MSAID2008 } \\
  \multicolumn{7}{c}{ epoch~1 } \\
  F115W  & 2022-11-04T10:30:46.369 & 2022-11-04T18:15:08.372 & 9985.2 & 59887.61260 & 0.13756 & 2561 \\
  F150W  & 2022-11-04T12:39:04.666 & 2022-11-04T19:39:36.143 & 9985.2 & 59887.68557 & 0.21053 & 2561 \\
  F200W  & 2022-11-04T14:47:12.210 & 2022-11-04T16:49:36.153 & 6699.7 & 59887.65863 & 0.18359 & 2561 \\
  F277W  & 2022-11-04T10:30:46.369 & 2022-11-04T12:17:25.481 & 5862.3 & 59887.47504 & 0.00000 & 2561 \\
  F356W  & 2022-11-04T12:39:04.666 & 2022-11-04T14:26:05.281 & 5862.3 & 59887.56430 & 0.08926 & 2561 \\
  F444W  & 2022-11-04T16:55:52.002 & 2022-11-04T19:39:36.207 & 8245.8 & 59887.76232 & 0.28728 & 2561 \\
  \multicolumn{7}{c}{ epoch~2 } \\
  F115W  & 2023-08-01T03:34:29.680 & 2023-08-01T11:26:14.683 & 8117.0  & 60157.30673 & 269.83169 & 2561 \\
  F150W  & 2023-08-01T11:32:08.972 & 2023-08-01T13:06:38.037 & 5024.8  & 60157.51351 & 270.03847 & 2561 \\
  F200W  & 2023-08-01T13:12:43.079 & 2023-08-01T14:47:33.583 & 5024.8  & 60157.58337 & 270.10833 & 2561 \\
  F277W  & 2023-08-01T03:34:29.616 & 2023-08-01T11:26:14.619 & 8117.0  & 60157.30673 & 269.83169 & 2561 \\
  F356W  & 2023-08-01T11:32:08.972 & 2023-08-01T13:06:37.973 & 5024.8  & 60157.51351 & 270.03847 & 2561 \\
  F444W  & 2023-08-01T13:12:43.079 & 2023-08-01T14:47:33.583 & 5024.8  & 60157.58337 & 270.10833 & 2561 \\
  \multicolumn{7}{c}{ epoch~3 } \\
  F115W  & 2023-10-26T01:47:35.574 & 2023-10-26T04:44:02.014 & 10049.6 & 60243.13598 & 355.66094 & 3990 \\
  F200W  & 2023-10-29T16:29:31.190 & 2023-10-29T19:25:57.629 & 10049.6 & 60246.74843 & 359.27339 & 3990 \\
  F277W  & 2023-10-25T21:13:27.654 & 2023-10-26T00:09:54.094 & 10049.6 & 60242.94561 & 355.47057 & 3990 \\
  F356W  & 2023-10-26T01:47:35.574 & 2023-10-26T04:44:02.014 & 10049.6 & 60243.13598 & 355.66094 & 3990 \\
  F444W  & 2023-10-29T16:29:31.190 & 2023-10-29T19:25:57.629 & 10049.6 & 60246.74843 & 359.27339 & 3990 \\
  \multicolumn{7}{c}{ epoch~4 } \\
  F356W  & 2023-12-01T18:00:33.083 & 2023-12-10T11:17:31.518 & 2104.4  & 60284.11047 & 396.63543 & 3516 \\\hline
  \multicolumn{7}{c}{ MSAID4286 } \\
  \multicolumn{7}{c}{ epoch~1 } \\
  F115W  & 2022-10-20T12:32:11.867 & 2022-10-20T13:12:06.239 &  2104.4 & 59872.53622 & 0.00118 & 2756 \\
  F150W  & 2022-10-20T13:16:45.337 & 2022-10-20T13:56:18.269 &  2104.4 & 59872.56703 & 0.03199 & 2756 \\
  F200W  & 2022-10-20T14:01:08.054 & 2022-10-20T14:40:40.986 &  2104.4 & 59872.59785 & 0.06281 & 2756 \\
  F277W  & 2022-10-20T14:01:08.118 & 2022-10-20T14:40:40.986 &  2104.4 & 59872.59785 & 0.06281 & 2756 \\
  F356W  & 2022-10-20T13:16:45.337 & 2022-10-20T13:56:18.205 &  2104.4 & 59872.56703 & 0.03199 & 2756 \\
  F444W  & 2022-10-20T12:32:11.931 & 2022-10-20T13:12:06.239 &  1578.3 & 59872.53504 & 0.00000 & 2756 \\
  \multicolumn{7}{c}{ epoch~2 } \\
  F115W  & 2022-12-06T09:09:17.444 & 2022-12-06T09:48:50.378 &  2104.4 & 59919.39518 & 46.86014 & 2756 \\
  F150W  & 2022-12-06T09:53:07.971 & 2022-12-06T10:32:40.903 &  2104.4 & 59919.42563 & 46.89059 & 2756 \\
  F200W  & 2022-12-06T10:36:58.496 & 2022-12-06T11:16:31.428 &  2104.4 & 59919.45608 & 46.92104 & 2756 \\
  F277W  & 2022-12-06T10:36:58.560 & 2022-12-06T11:16:31.428 &  2104.4 & 59919.45608 & 46.92104 & 2756 \\
  F356W  & 2022-12-06T09:53:07.971 & 2022-12-06T10:32:40.903 &  2104.4 & 59919.42563 & 46.89059 & 2756 \\
  F444W  & 2022-12-06T09:09:17.509 & 2022-12-06T09:48:50.378 &  2104.4 & 59919.39518 & 46.86014 & 2756 \\ 
  \multicolumn{7}{c}{ epoch~3 } \\
  F115W  & 2023-10-26T01:47:35.510 & 2023-10-26T04:44:02.014 &  10049.6 & 60243.13598 & 370.60094 & 3990 \\
  F200W  & 2023-10-29T16:29:31.062 & 2023-10-29T19:25:57.629 &  10049.6 & 60246.74843 & 374.21339 & 3990 \\
  F277W  & 2023-10-25T21:13:27.590 & 2023-10-26T00:09:54.094 &  10049.6 & 60242.94561 & 370.41057 & 3990 \\
  F356W  & 2023-10-26T01:47:35.574 & 2023-10-26T04:44:02.014 &  10049.6 & 60243.13598 & 370.60094 & 3990 \\
  F444W  & 2023-10-29T16:29:31.062 & 2023-10-29T19:25:57.629 &  10049.6 & 60246.74843 & 374.21339 & 3990 \\ 
  \multicolumn{7}{c}{ epoch~4 } \\
  F356W  & 2023-12-05T21:01:03.274 & 2023-12-10T16:20:15.148 &  1052.2 & 60286.27823 & 413.74319 & 3516 \\\hline  
  \multicolumn{7}{c}{ MSAID38108 } \\
  \multicolumn{7}{c}{ epoch~1 } \\
  F115W  & 2022-06-29T02:18:38.805 & 2022-06-29T06:26:18.493 & 11767.5 & 59759.18781 & 0.08811 & 1324 \\
  F150W  & 2022-06-29T06:32:12.730 & 2022-06-29T08:42:18.410 & 6195.1  & 59759.32500 & 0.22530 & 1324 \\
  F200W  & 2022-06-29T08:47:08.260 & 2022-06-29T10:40:46.115 & 5572.4  & 59759.41450 & 0.31480 & 1324 \\
  F277W  & 2022-06-29T08:47:08.324 & 2022-06-29T10:40:46.179 & 5572.4  & 59759.41450 & 0.31480 & 1324 \\
  F356W  & 2022-06-29T06:32:12.794 & 2022-06-29T08:42:18.410 & 6195.1  & 59759.32500 & 0.22530 & 1324 \\
  F444W  & 2022-06-28T22:04:43.300 & 2022-06-29T06:26:18.493 & 23535.0 & 59759.09970 & 0.00000 & 1324 \\
  \multicolumn{7}{c}{ epoch~2 } \\
  F115W  & 2022-11-02T08:27:04.089 & 2022-11-11T00:08:25.433 & 27314.3 & 59889.11297 & 130.01327 & 1324, 2561 \\
  F150W  & 2022-11-02T10:37:52.658 & 2022-11-11T02:48:13.392 & 19068.5 & 59887.90799 & 128.80829 & 1324, 2561 \\
  F200W  & 2022-11-02T12:48:19.786 & 2022-11-11T05:26:35.462 & 14945.6 & 59889.26121 & 130.16151 & 1324, 2561 \\
  F277W  & 2022-11-02T08:27:04.153 & 2022-11-11T02:48:13.392 & 14945.6 & 59889.11049 & 130.01079 & 1324, 2561 \\
  F356W  & 2022-11-02T10:37:52.722 & 2022-11-11T05:26:35.462 & 14945.6 & 59889.20954 & 130.10984 & 1324, 2561 \\
  F444W  & 2022-11-02T14:59:08.410 & 2022-11-11T00:08:25.433 & 41229.2 & 59890.54602 & 131.44632 & 1324, 2561 \\
  \multicolumn{7}{c}{ epoch~3 } \\
  F115W  & 2023-07-07T10:20:51.741 & 2023-07-07T14:29:14.437 & 11767.5 & 60132.52304 & 373.42334 & 1324 \\
  F277W  & 2023-07-07T10:20:51.805 & 2023-07-07T11:58:44.814 & 4949.7  & 60132.46138 & 373.36168 & 1324 \\
  F356W  & 2023-07-07T06:06:56.383 & 2023-07-07T07:44:27.953 & 4949.7  & 60132.28484 & 373.18514 & 1324 \\
  F444W  & 2023-07-07T07:50:32.953 & 2023-07-07T14:29:14.437 & 13635.7 & 60132.46278 & 373.36308 & 1324 \\
  \multicolumn{7}{c}{ epoch~4 } \\
  F356W  & 2023-11-27T10:40:12.083 & 2023-12-10T06:17:43.272 & 3156.6 & 60281.85386 & 522.75416 & 3516 \\  
\enddata
\tablenotetext{}{Start and end times are the UTC times at the start and end of exposure ({\tt DATE-BEG} and {\tt DATE-END} recorded in the fits header of the mosaic images. Exposure time is the effective exposure time ({\tt EFFEXPTM}). MJD-mid is the exposure mid-point in MJD ({\tt EXPMID}), and $\Delta t_{\text{obs}}$ is the observer-frame time separation relative to the minimum MJD-mid. Program IDs: ERS~1324 (GLASS; PI: T.~L.~Treu), GO~2561 (UNCOVER; PI: I.~Labbe), DD~2756 (PI: W.~Chen), GO~3516 (PI: J.~Matthee), and GO~3990 (BEACON; PI: T.~Morishita). }
\end{deluxetable*}

Table~\ref{tbl:observations} summarizes the log of the {\it JWST}/NIRCam wide-band observations with which any of the five broad H$\alpha$ emitters are imaged, divided by `epochs' (Table~\ref{tbl:summary_data}) for each object.

Since GLASS~160133 and GLASS~150029 are located close to each other at a separation of $11''.68$, they were analyzed on the same mosaic images.
GLASS~160133 and GLASS~150029 were imaged twice in November 2022, August 2023 with the temporal separation of $\sim 272$~days in the observer-frame by {\it JWST}/NIRCam with the wide-band filters in the UNCOVER program \citep[Program ID: GO 2561, PI: I.~Labbe;][]{bez22}; see Table~\ref{tbl:observations}.
Additional NIRCam data in the F356W filter were obtained in December 2023 \citep[Program ID: GO 3516, PI: J.~Matthee;][]{nai24}.

MSAID2008, MSAID4286, and MSAID38108 were observed with the NIRCam wide-bands in several GO and DD programs.
MSAID2008 was simultaneously imaged as GLASS~160133 and GLASS~150029, and additionally observed in October 2023 in the BEACON program \citep[Program ID: GO~3990, PI: T.~Morishita;][]{mor24}.
MSAID4286 was observed in the program DD~2756 (PI: W.~Chen), GO~3516 (PI: J.~Matthee), and GO~3990 \citep[see e.g.,][]{par23,sue24}.
MSAID38108 was observed in the GLASS program \citep[Program ID: ERS~1324, PI: T.~L.~Treu;][]{tre22}, UNCOVER program (GO~2561), and GO~3516.

After a visual inspection of the {\tt cal} images, the following exposures, which seem to be affected by guiding failure and cosmic-ray hitting, were removed from the analysis: jw02561001003\_02101\_00008 (GLASS~160133, GLASS~150029, MSAID2008; F115W, F277W), jw02561001003\_04101\_00008 (GLASS~160133, GLASS~150029, MSAID2008; F150W, F356W), and jw02756003001\_03101\_00003 (MSAID4286; F444W).

\section{Multi-band data likelihood of the DRW model}
\label{sec:data_likelihood}

Here, we present the irregularly sampled multi-band light curve data likelihood of the DRW model extending the approach of \cite{ryb92} \citep[see also][]{koz10,zu11}.

The observation equation of an object's apparent magnitude $y$ measured in a photometric band of a pivot wavelength $\lambda_{a}$ ($a=1,2,\dots$, $N$) in the rest-frame at a given epoch $t_{i}$ ($i=1,2,\dots$, $M$) in the rest-frame is given as:
\begin{eqnarray}
y(\lambda_{a}, t_{i}) &=& s(\lambda_{a}, t_{i}) + n(\lambda_{a}, t_{i}) + q(\lambda_{a}),
\label{eqb:observation_equation}
\end{eqnarray}
where $s(\lambda_{a}, t_{i})$ is the latent signal, and $n(\lambda_{a}, t_{i})$ is the heteroskedastic Gaussian measurement noise.
$q(\lambda_{a})$ is a wavelength-dependent time-independent magnitude to represent the mean of the light curve (e.g., consisting of the non-variable host galaxy flux plus average AGN flux).
In the case of the NIRCam data we use in the main text, 
$(\lambda_{1}, \lambda_{2}, \lambda_{3}, \lambda_{4}, \lambda_{5}, \lambda_{6}) = (1.154, 1.501, 1.990, 2.786, 3.563, 4.421)/(1+z)$ in units of $\mu$m for F115W, F150W, F200W, F277W, F356W, and F444W, respectively, 
and the temporal separation between the two epochs $\Delta t_{ij} = t_{i} - t_{j}$ ($i, j = 1,2,\dots$, $M$) corresponds to the difference of MJD-mid in Table~\ref{tbl:observations} divided by $1+z$.

The measurement noises are mutually independent random variables drawn from the Gaussian distribution of the dispersion $\sigma_{n}$:
\begin{eqnarray}
n(\lambda_{a}, t_{i}) = G\left[\sigma_{n}(\lambda_{a}, t_{i})^2\right]
\end{eqnarray}
where $G(x^2)$ is a Gaussian deviate of the dispersion $x$, and its covariance is:
\begin{eqnarray}
\langle n(\lambda_{a}, t_{i})n(\lambda_{b}, t_{j})\rangle &=& \langle n(\lambda_{b}, t_{j})n(\lambda_{a}, t_{i})\rangle = \delta_{ab}\delta_{ij}\sigma_{n}(\lambda_{a}, t_{i})^2,
\label{eqn:noise_covariance}
\end{eqnarray}
where $\delta$ is the Kronecker delta.

We assume the DRW process as a time-series model describing the latent signals $s(\lambda_{a}, t_{i})$.
The relationship between signals at $t_{i}$ and $t_{j}$ in a single band $\lambda_{a}$ is defined as \citep[e.g.,][]{koz10}:
\begin{eqnarray}
s(\lambda_{a}, t_{i}) &=& s(\lambda_{a}, t_{j})e^{-\frac{|\Delta t_{ij}|}{\tau_{\text{d}}}} + G\left[\sigma_{\text{d}}(\lambda_{a})^2\left(1-e^{-\frac{2|\Delta t_{ij}|}{\tau_{\text{d}}}}\right)\right],
\label{eqn:drw}
\end{eqnarray}
where $\Delta t_{ij} = t_{i} - t_{j}$.
$\sigma_{\text{d}}(\lambda_{i})$ is the wavelength-dependent asymptotic variability amplitude, and $\tau_{\text{d}}$ is the decorrelation time scale of the DRW model, respectively.
The structure function (SF) of the DRW model is $\text{SF}(\lambda_{a}, \Delta t_{ij}) \equiv \sqrt{\langle (s(\lambda_{a},t_{i})-s(\lambda_{a},t_{j}))^2 \rangle} = \sqrt{2}\sigma_{\text{d}}(\lambda_{a})\sqrt{1-\exp\left(-|\Delta t_{ij}|/\tau_{\text{d}}\right)}$, from which the asymptotic SF is defined as $\text{SF}_{\infty}(\lambda_{a}) \equiv \sqrt{2}\sigma_{\text{d}}(\lambda_{a})$ \citep{mac10}.
This SF implies that the DRW process behaves as a random walk on short time scales ($\text{SF}(\lambda_{a}, \Delta t_{ij}) \approx \text{SF}_{\infty}(\lambda_{a})\sqrt{|\Delta t_{ij}|/\tau_{\text{d}}}$ at $\Delta t_{ij}<\tau_{\text{d}}$) and asymptotically approaches a white noise with a finite amplitude on long time scales ($\text{SF}(\lambda_{a}, \Delta t_{ij}) \approx \text{SF}_{\infty}(\lambda_{a})$ at $\Delta t_{ij}>\tau_{\text{d}}$).

Here, we further assume that the multi-band signals are perfectly correlated at zero lag, which is an adequate approximation for the AGN UV-optical accretion disk continuum variability.
That is, $\tau_{\text{d}}$ is assumed to be wavelength-independent, and the signal covariance between $s(\lambda_{a}, t_{i})$ and $s(\lambda_{b}, t_{j})$ can be expressed as:
\begin{eqnarray}
\langle s(\lambda_{a}, t_{i})s(\lambda_{b}, t_{j})\rangle &=& \langle s(\lambda_{b}, t_{j})s(\lambda_{a}, t_{i})\rangle = \sigma_{\text{d}}(\lambda_{a})\sigma_{\text{d}}(\lambda_{b}) e^{-\frac{|\Delta t_{ij}|}{\tau_{\text{d}}}}.
\label{eqn:signal_covariance}
\end{eqnarray}

Consider a dataset that is irregularly sampled in terms of both wavelengths and epochs. 
For each epoch $t_{i}$ ($i=1,\dots,M$), there are multi-band data sampled at $N_{t_{i}}$ wavelengths $(\lambda_{(t_{i}, 1)}, \lambda_{(t_{i}, 2)}, \dots, \lambda_{(t_{i}, N_{t_{i}})}) \in \{\lambda_{1}, \lambda_{2}, \dots, \lambda_{N}\}^{N_{t_{i}}}$.
A full data vector $\vec{y}$ is defined as being composed of measurements from all $M$ epochs, by arranging $y(\lambda_{a}, t_{i})$ in Equation~\ref{eqb:observation_equation} into a vector of $d \equiv \sum_{i=1}^{M} N_{t_{i}}$ dimensions as 
\begin{eqnarray}
\vec{y} &=& \left( y(\lambda_{(t_{1}, 1)}, t_{1}), y(\lambda_{(t_{1}, 2)}, t_{1}), \dots, y(\lambda_{(t_{1}, N_{t_{1}})}, t_{1}), y(\lambda_{(t_{2}, 1)}, t_{2}), \dots, y(\lambda_{(t_{M}, N_{t_{M}}-1)}, t_{M}), y(\lambda_{(t_{M}, N_{t_{M}})}, t_{M}) \right)^{T}.
\label{eqn:data_vector}
\end{eqnarray}
Likewise, the signal vector $\vec{s}$, noise vector $\vec{n}$, and constant flux vector $\vec{q}$ are defined from the following vector equation:
\begin{eqnarray}
\vec{y} &\equiv& \vec{s} + \vec{n} + L\vec{q},
\end{eqnarray}
where $\vec{q} \equiv (q(\lambda_{1}), \dots, q(\lambda_{N}))^{T}$ is the $N$-dimensional constant vector.
$L$ is the $d\times N$ matrix, which has entries of $(1, 0, \dots, 0)$ for the $\lambda = \lambda_{1}$ data points, $(0, 1, \dots, 0)$ for the $\lambda = \lambda_{2}$ data points, and so forth \citep[e.g.,][]{ryb92,zu11,zu16}.
The $d\times d$ covariance matrix of $\vec{s}$ and $\vec{n}$ are respectively denoted as $S$ and $N$, where $S$ is the symmetric dense matrix (Equation~\ref{eqn:signal_covariance}) and $N$ is the diagonal matrix (Equation~\ref{eqn:noise_covariance}).

In terms of the Bayesian probability, assuming the Gaussian process means to adopt a prior distribution of $\vec{s}$ given the set of hyperparameters $\vec{\theta} \equiv \{\tau_{\text{d}}, \sigma_{\text{d}}(\lambda_{1}), \dots, \sigma_{\text{d}}(\lambda_{N})\}$ as \citep[e.g.,][]{ryb92,koz10}:
\begin{eqnarray}
p(\vec{s}|\vec{\theta}) = \frac{1}{\sqrt{(2\pi)^d |S|}}\exp\left( -\frac{\vec{s}^{T}S^{-1}\vec{s}}{2} \right).
\end{eqnarray}
According to the observation equation, $\vec{y}$ is conditionally independent of the hyperparameters given the latent variable $\vec{s}$ \citep[][]{ras06}, and the probability of $\vec{y}$ given $\vec{s}$ and $\vec{q}$, i.e., the data likelihood, is defined as
\begin{eqnarray}
p(\vec{y}|\vec{s},\vec{q}) = \frac{1}{\sqrt{(2\pi)^d |N|}}\exp\left( -\frac{\left(\vec{y}-L\vec{q}-\vec{s}\right)^{T}N^{-1}\left(\vec{y}-L\vec{q}-\vec{s}\right)}{2} \right).
\end{eqnarray}
An integral of the likelihood $p(\vec{y}|\vec{s},\vec{q})$ multiplied by $p(\vec{s}|\vec{\theta})$ and a vague prior on $\vec{q}$ ($p(\vec{q}) \propto 1$) over $\vec{s}$ and $\vec{q}$ defines a marginal data likelihood as \citep[][]{ras06,koz10,zu11}:
\begin{eqnarray}
p(\vec{y}|\vec{\theta}) &\propto& \frac{1}{\sqrt{(2\pi)^{d/2}|C||C_{q}^{-1}|}}\exp\left( -\frac{\vec{y}^{T}C_{\perp}^{-1}\vec{y}}{2} \right),
\label{eqn:datalikelihood}
\end{eqnarray}
where $C \equiv S + N$ is the data covariance matrix, $C_{q} \equiv (L^{T}C^{-1}L)^{-1}$, and $C_{\perp}^{-1} \equiv C^{-1}-C^{-1}LC_{q}L^{T}C^{-1}$.
In a special case of the single-band two-epoch data ($N=1, M=2$, $\vec{\theta}=\{\tau_{\text{d}}, \sigma_{\text{d}}(\lambda_{1})\}$), the data likelihood reduces to:
\begin{eqnarray}
p(\vec{y}|\vec{\theta}) &\propto& \frac{1}{\sqrt{(2\pi)|C||C_{q}^{-1}|}}\exp\left( -\frac{\left(y(\lambda_{1}, t_{2})-y(\lambda_{1}, t_{1})\right)^2}{2|C||C_{q}^{-1}|} \right),
\label{eqn:datalikelihood_singleband}
\end{eqnarray}
which is the Gaussian with the variance given as the sum of the signal and noise variance:
\begin{eqnarray}
|C||C_{q}^{-1}| = 2\sigma_{\text{d}}(\lambda_{1})^2\left(1-e^{-\frac{|\Delta t|}{\tau_{\text{d}}}}\right) + \sigma_{n}(\lambda_{1},t_{1})^2 + \sigma_{n}(\lambda_{1},t_{2})^2 = \text{SF}(\lambda_{1}, \Delta t)^2 + \sigma_{n}(\lambda_{1},t_{1})^2 + \sigma_{n}(\lambda_{1},t_{2})^2.
\end{eqnarray}

The signal covariance matrix $S$ is singular if a part of the multi-band data is obtained at the same epochs because the multi-band data at the same epoch without measurement noise are not independent of each other (Equation~\ref{eqn:signal_covariance}).
Meanwhile, even in this case, the data covariance matrix $C$ is positive definite since $N$ works as the regularization matrix \citep{ras06}; thus, Equation~\ref{eqn:datalikelihood} is well defined.
Due to the symmetry of the covariance matrix, the matrix calculation in Equation~\ref{eqn:datalikelihood} can be performed using the Cholesky factorization of $C$ and $C_{q}^{-1}$.
Note that Equation~\ref{eqn:datalikelihood} is invariant under displacements of $\vec{y}$ as $\vec{y'} = \vec{y} + L\vec{p}$ where $\vec{p}$ is an arbitrary constant $N$-dimensional vector \citep[$\vec{y'}^{T}C_{\perp}^{-1}\vec{y'} = \vec{y}^{T}C_{\perp}^{-1}\vec{y}$;][]{ryb92}, which means that the systematic offsets in the absolute magnitude calibration (including corrections for the extinction, finite aperture effect, and emission line contributions) do not influence the data likelihood $p(\vec{y}|\vec{\theta})$.


\bibliography{jwst_agn_var}{}
\bibliographystyle{aasjournal}



\end{document}